\documentclass[showpacs,amssymb,preprint,preprintnumbers,nofootinbib,superscriptaddress]{revtex4}
\usepackage{amsmath}
\usepackage{graphicx}
\graphicspath{{Figures/}}
\usepackage{latexsym}
\usepackage{amsfonts}
\usepackage{url,hyperref}
\usepackage{bm}
\usepackage{textcomp}
\usepackage{color}
\usepackage{bbm}
\usepackage{slashed}
\usepackage{caption}
\usepackage{epstopdf}
\usepackage{subcaption}
\usepackage{placeins}
\usepackage{color}

\newcommand{\beq}{\begin{equation}}
\newcommand{\eeq}{\end{equation}}
\def\bea{\begin{eqnarray}}
\def\eea{\end{eqnarray}}

\begin{document}

\title{Quasinormal modes of a massless Dirac field in the dRGT massive gravity}

\author{Pitayuth Wongjun \footnote{Email: pitbaa@gmail.com}}
\affiliation{The institute for fundamental study, Naresuan University, Phitsanulok 65000, Thailand}
\affiliation{Thailand Center of Excellence in Physics, Ministry of Higher Education, Science, Research and Innovation, 328 Si Ayutthaya Road, Bangkok 10400, Thailand}

\author{Chun-Hung Chen \footnote{Email: chun-hungc@nu.ac.th}}
\affiliation{The institute for fundamental study, Naresuan University, Phitsanulok 65000, Thailand}

\author{Ratchaphat Nakarachinda \footnote{Email: tahpahctar\_net@hotmail.com}}
\affiliation{The institute for fundamental study, Naresuan University, Phitsanulok 65000, Thailand}
\affiliation{Thailand Center of Excellence in Physics, Ministry of Higher Education, Science, Research and Innovation, 328 Si Ayutthaya Road, Bangkok 10400, Thailand}

\begin{abstract}
The quasinormal modes of a massless Dirac field in the de Rham-Gabadadze-Tolley (dRGT) massive gravity theory with asymptotically de Sitter spacetime are investigated using the Wentzel-Kramers-Brillouin (WKB) approximation. The effective potential for the massless Dirac field due to the dRGT black hole is derived. It is found that the shape of the potential depends crucially on the structure of the graviton mass and the behavior of the quasinormal modes is controlled by the graviton mass parameters. Higher potentials give stronger damping of the quasinormal modes. We compare our results to the Schwarzschild-de Sitter case. Our numerical calculations are checked using Pad$\acute{e}$ approximation and found that the quasinormal mode frequencies converge to ones with reasonable accuracy.

\end{abstract}
\maketitle

\section{Introduction}

General Relativity (GR) is one of pillars of modern physics. It has been confirmed by many kinds of observation, e.g., the precession of Mercury's orbit \cite{Clemenc:1947}, gravitational time dilation \cite{Schwartz:1977,Uggerhoj:2016} and, recently, gravitational waves \cite{Abbott:2016blz}. According to observations, the universe is expanded with acceleration at the present \cite{Riess:1998cb,Perlmutter:1998np}. Including a cosmological constant to GR is one of the possible ways to explain this expansion. Even though GR with a cosmological constant can provide a description able to satisfy current observations, the physical origin of the cosmological constant still has not been conclusively explained. There are several ways to explain such a phenomenon of the universe instead of using the cosmological constant. One of such ways is trying to modify gravity at large scale. Massive gravity theory is a kind of such modifications in which a mass term is given.

A linear theory of massive gravity, Fierz-Pauli massive gravity, was proposed in 1939 as the theory of the massive spin-2 field \cite{Fierz:1939ix}. Unfortunately, this theory encounters the van Dam-Veltman-Zakharov (vDVZ) discontinuity in the massless limit \cite{vanDam:1970ab,Zakharov:1970cd}. In other words, the Fierz-Pauli theory cannot be reduced to a linearlized version of GR. It was suggested by Vainshtein \cite{Vainshtein:1972sx} that nonlinear interaction terms should be included into the theory to get rid of vDVZ discontinuity. One viable model of nonlinear massive gravity theories was proposed in 2010 by de Rham, Gabadadze and Tolley, and is called the dRGT massive gravity \cite{deRham:2010ik,deRham:2010kj}. One of the key points of this massive gravity theory is that Struckelberg fields are introduced via the reference/fiducial metric to restore diffeomorphism invariance. By using a Minkowski type fiducial metric, it is found that the dRGT massive gravity theory does not admit flat Friedmann-Lema\^{i}tre-Robertson-Walker (FLRW) solutions, so that it is not easy to obtain suitable models to provide the present-day acceleration of the universe \cite{DAmico:2011eto,Gumrukcuoglu:2011ew}. Further studies of the dRGT massive gravity have been investigated in order to provide the acceleration phase of the universe, for example, considering more general forms of the fiducial metric \cite{Fasiello:2012rw,Langlois:2012hk,Langlois:2013cya,Gumrukcuoglu:2011zh,Chullaphan:2015ija}, including more degrees of freedom, such as a scalar field \cite{Huang:2012pe,DAmico:2012hia,DeFelice:2013tsa,DeFelice:2013dua,DeFelice:2017wel,DeFelice:2017rli,Hinterbichler:2013dv,Gabadadze:2012tr,Tannukij:2015wmn,Nakarachinda:2017oyc}, and promoting the fiducial metric to a dynamical field \cite{HassanRosen2012}. Nice reviews on the massive gravity theory are also found \cite{Hinterbichler, deRham:2014zqa}.

Since GR is modified by graviton mass at large scale, it is possible that the local gravity may obtain some modification due to the graviton mass. As a result, a spherically symmetric solution in dRGT massive gravity  has been investigated in order to examine effects of such modification at local scale \cite{Koyama:2011yg,Koyama:2011xz,Nieuwenhuizen:2011sq,Tasinato:2013rza,Vegh:2013sk,BHsoln1,BHsoln2,BHsoln3}. By using this kind of solution, some signatures of astronomical objects in the dRGT massive gravity have been explored, e.g., in white dwarfs \cite{EslamPanah:2018evk}, neutron stars \cite{Hendi:2017ibm}, the rotation curves of galaxies \cite{Panpanich:2018cxo}, gravitational lensing \cite{Panpanich:2019mll} and the mass-radius ratio bound for compact objects \cite{Kareeso:2018qea}. Moreover, black hole solutions were also found \cite{Berezhiani:2011mt,Brito:2013xaa,Volkov:2013roa,Cai:2012db,Babichev:2014fka,Babichev:2015xha,Hu:2016hpm} and their thermodynamical properties have been investigated in \cite{Cai:2014znn,Ghosh:2015cva,Adams:2014vza,Xu:2015rfa,Capela:2011mh,Hu:2016mym,Zou:2016sab,Hendi:2017arn,Hendi:2017bys,EslamPanah:2016pgc,Hendi:2016hbe,Hendi:2016uni,Hendi:2016yof,Arraut:2014uza,Arraut:2014iba}. Besides the spherically symmetric solution, the cylindrical one has also investigated in \cite{Tannukij:2017jtn,Ponglertsakul:2018smo,Boonserm:2019mon,Ghosh:2019eoo}.

The linear perturbations in GR around various kinds of black hole solutions such as Schwarzschild, Reissiner-Nordstr\"{o}m and Kerr are well investigated \cite{Chand1083}. For spherically symmetric spacetime, the general form of the equations of motion for gravitational perturbations, as well as their instabilities, have been investigated \cite{Kodama:2003jz,Ishibashi:2003ap,Kodama:2003kk}. It was found that the governing equations in linear perturbation regime are closely related to the equations of a field in curved spacetime around the black holes. Thus, in order to study the instabilities and properties of black holes, it is possible to analyze the behavior of a test field nearby, dictated as a gravitational interaction of the field itself. As a result, various kinds of the fields have been investigated, for example, scalar field \cite{BCS2009}, Dirac field \cite{CCDW2009}, vector field \cite{chm2001} and spin-3/2 gravitino field \cite{ccch2019}. As modified gravity theories have been developed, the black hole solutions in such the theories were found. In order to explore the instabilities and the dynamical properties of black holes in modified gravity, one can analyze the evolution of the field around the black hole. Up to our knowledge, the scalar field around the black hole in the dRGT massive gravity had been examined \cite{Burikham:2017gdm,Boonserm:2017qcq,Boonserm}, the Dirac field has not been investigated yet. Hence, we focus on this investigation in the present work through the Quasinormal modes (QNMs).

QNMs are solutions of the wave equation with specific boundary conditions and contain the complex frequency. Analyzing the QNMs is one way to investigate the dynamical properties of a black hole. For example, the ringdown frequency profile of black hole merging is characterized by the QNMs (see, e.g., \cite{Gundlach:1993tn,Gundlach:1993tp} and \cite{Kokkotas:1999bd} for a review). This is interesting, since a new generation of gravitational wave detectors may be able to detect some signatures of the QNMs, and so might provide some hint to construct a modified gravity theory. QNMs are also great interesting in the context of the Anti-de Sitter/Conformal field theory (AdS/CFT) correspondence \cite{Horowitz:1999jd}, as well as in the context of the thermodynamic properties of black holes in loop quantum gravity (see \cite{Dreyer:2002vy} and \cite{Cardoso:2003pj} for reviews). Because of the nature of the dRGT black hole solution, which is asymptotically Anti-de Sitter (AdS)/de Sitter (dS), it is worthwhile to investigate the QNMs of the dRGT black hole and this is the main aim of this work.

It is a common knowledge that the scalar perturbations around spherically symmetric black holes are governed by the Regge-Wheeler equation. For the perturbation of the Dirac field, the master equations were obtained \cite{Cho:2003qe,Cho:2005yc}. It is also found that the mathematical form of these master equations are different to ones of the bosonic cases. Furthermore, no spherically symmetric black hole solutions for the Einstein-Dirac-Maxwell system and the absence of a periodic static orbit for a Dirac particle around a black hole were proved by Finster and collaborators \cite{fsy1999,fsy2000,fksy2000}. However, according to Hawking radiation, it is possible to have radiation around the black hole as  quantum fluctuations of Dirac particles. This implies that the scattering properties of a Dirac particle, or a “Dirac cloud”, around a black hole are interesting to investigate. 

Various methods were established to compute QNMs, e.g., the Poshl-Teller method \cite{Ferrari:1984zz}, the asymptotic iteration method \cite{Cho:2009cj,Cho:2011sf} and the WKB method \cite{SchutzWill,iyewil}. A review of QNMs in various kinds of black holes can be found in \cite{BCS2009,KZ2011}. In this work, we evaluated the QNMs of the Dirac field surrounding the dRGT black holes by using WKB and revised WKB method with Pad$\acute{e}$ approximation \cite{KZZ2019}. From a cosmological point of view, the universe is expected to be asymptotically dS on large scale. In particular, the QNMs in this kind of spacetime have been intensively investigated \cite{Mellor:1989ac,Moss:2001ga,Cardoso:2003sw,Molina:2003ff,Suneeta:2003bj,MaassenvandenBrink:2003yq,Choudhury:2003wd,Jing:2005bh,Ghosh:2005aq,LopezOrtega:2006my,Yoshida:2010zzb,Liu:2012zl,Zhang:2014xha,Tangphati:2018jdx} including for Dirac fields \cite{Zhi2003,Jing:2003wq,Wahlang:2017zvk}. Hence, we will focus on the asymptotically dS solution of the dRGT black hole. We analyze the Schr\"{o}dinger-like equation for Dirac perturbation with a particular form of the potentials due to the dRGT black hole. We find that the shape of the potentials depends crucially on the structure of the graviton mass and that the behavior of the QNMs is controlled by the graviton mass parameters. It is also found that the higher potential gives stronger damping of the QNMs. Lastly, we compare our results to the Schwarzschild-de Sitter case and find that the Dirac QNMs for the Schwarzschild-de Sitter black hole are located, approximately, in a part of parameter space from the dRGT black hole.

This paper is organized as follows. In Sec. \ref{sec:dRGT}, a brief review of the dRGT massive gravity theory and its black hole solution are discussed. Then, in Sec. \ref{sec:eff potential}, the effective potential in Schr\"{o}dinger-like equation is derived and analyzed. We show how the shape of the potential depends on the mass parameters of the theory. In Sec. \ref{sec:QNM}, the QNMs are computed using the WKB method. Moreover, the accuracy of the computations are checked using the Pad$\acute{e}$ approximation. Sec. \ref{sec:conclude} contains a summary of our main conclusions and discussion. The tables of QNM frequencies and the numerical precision are listed in Appendices A and B, respectively.


\section{dRGT massive gravity and black hole solution}\label{sec:dRGT}
 A massive gravity theory is a modified gravity theory that introduces a mass term into GR. One of the viable models of massive gravity is proposed by de Rham, Gabadaze and Tolley, called the dRGT massive gravity \cite{deRham:2010ik,deRham:2010kj}. The action for the dRGT massive gravity can be written as
\begin{eqnarray}\label{action}
 S = \int \text{d}^4x \sqrt{-g}\; \frac{1}{2} \left[ R +m_g^2\,\, {\cal U}(g, \phi^a)\right],
\end{eqnarray}
where $R$ is the Ricci scalar and ${\cal U}$ is the potential for the graviton. The letter is an additional part of gravitational sector with the parameter $m_g$ interpreted as the graviton mass. The potential ${\cal U}$ in four-dimensional spacetime is of the form
\begin{eqnarray}\label{potential}
	{\cal U}(g, \phi^a) = {\cal U}_2 + \alpha_3{\cal U}_3 +\alpha_4{\cal U}_4 ,
\end{eqnarray}
where $\alpha_3$ and $\alpha_4$ are dimensionless free parameters of the theory and each term of the potential ${\cal U}_2$, ${\cal U}_3$ and ${\cal U}_4$, can be further expressed as
\begin{eqnarray}
	{\cal U}_2&\equiv&[{\cal K}]^2-[{\cal K}^2] ,\\
	{\cal U}_3&\equiv&[{\cal K}]^3-3[{\cal K}][{\cal K}^2]+2[{\cal K}^3] ,\\
	{\cal U}_4&\equiv&[{\cal K}]^4-6[{\cal K}]^2[{\cal K}^2]+8[{\cal K}][{\cal
K}^3]+3[{\cal K}^2]^2-6[{\cal K}^4],
\end{eqnarray}
where
\begin{eqnarray}
	{\cal K}^\mu_{\,\,\,\nu}=\delta^\mu_\nu-\sqrt{g^{\mu\sigma}f_{ab}\partial_\sigma\phi^a\partial_\nu\phi^b}. \label{K-tensor}
\end{eqnarray}
Here the rectangular brackets denote traces, namely $[{\cal K}]={\cal K}^\mu_{\,\,\,\mu}$ and $[{\cal K}^n]=({\cal K}^n)^\mu_{\,\,\,\mu}$. From the above expression, one can see that there exists another metric $f_{ab}$ called reference (or fiducial) metric. The four scalar fields $\phi^a$, called St\"uckelberg fields, are introduced in order to restore the general covariance of the theory.

By varying the action with respect to metric $g_{\mu\nu}$, the equations of motion interpreted as modified Einstein field equations are obtained as
\begin{eqnarray}
	G_{\mu\nu} +m_g^2 X_{\mu\nu} = 0. \label{modEFE}
\end{eqnarray}
The tensor $X_{\mu\nu}$ can be interpreted as the effective energy-momentum tensor. It is straightforwardly obtained by varying the potential term ${\cal U}$ with respect to $g_{\mu\nu}$
\begin{eqnarray}
	X_{\mu\nu} &=& {\cal K}_ {\mu\nu} -{\cal K}g_ {\mu\nu} -\alpha\left({\cal K}^2_{\mu\nu}-{\cal K}{\cal K}_{\mu\nu} +\frac{{\cal U}_2}{2}g_{\mu\nu}\right) 
 	+3\beta\left( {\cal K}^3_{\mu\nu} -{\cal K}{\cal K}^2_{\mu\nu} +\frac{{\cal U}_2}{2}{\cal K}_{\mu\nu} - \frac{{\cal U}_3}{6}g_{\mu\nu} \right), \,\,\,\,\,\,\label{effemt}\nonumber\\
\end{eqnarray}
where we have reparameterized the model parameters as follows
\begin{eqnarray}\label{alphabeta}
	\alpha_3 = \frac{\alpha-1}{3}~,~~~\alpha_4=\frac{\beta}{4}+\frac{1-\alpha}{12}.
\end{eqnarray}
Since the potential terms are covariantly constructed, the tensor $X_{\mu\nu}$ obeys the covariant divergence as follows
\begin{eqnarray}\label{BiEoM}
	\nabla^\mu X_{\mu\nu} = 0,
\end{eqnarray}
where $\nabla^\mu$ denotes the covariant derivative which is compatible with $g_{\mu\nu}$. Note that this constraint equation is also obtained by varying the action with respect to the fiducial metric, which also satisfies the Bianchi identities.

In order to solve the field equation \eqref{modEFE}, one may need to choose the form of the fiducial metric. Note that the form of the fiducial metric will provide the form of the physical metric. Considering this, it is convenient to choose the form of the fiducial metric as \cite{Ghosh:2015cva}
\begin{eqnarray}\label{fiducial metric}
f_{\mu\nu}=\text{diag}(0,0,h^2  ,h^2 \sin^2\theta), \label{fmetric}
\end{eqnarray}
where $h$ is a constant. By using this form of the fiducial metric, one of the static and spherically symmetric solutions of the physical metric can be obtained as
\begin{eqnarray}
	\text{d}s^2=-f(r)\text{d}t^2+f^{-1}(r)\text{d}r^2+r^2\text{d}\Omega^2,
\end{eqnarray}
with
\begin{eqnarray}\label{f sol}
	f(r)=1-\frac{2M}{r}+\frac{\Lambda}{3}r^2+\gamma r+\zeta,
\end{eqnarray}
where $M$ is the mass of black hole and other parameters are defined as follows
\begin{eqnarray}
	\Lambda&=&3m_g^2(1+\alpha+\beta),\\
	\gamma&=&-hm_g^2(1+2\alpha+3\beta),\\
	\zeta&=&h^2m_g^2(\alpha+3\beta).
\end{eqnarray}
Note that detailed calculation to obtain this solution can be found in \cite{Ghosh:2015cva}. This solution contains various signatures of other well-known black hole solutions found in the literature. By setting $m_{g}=0$, the Schwarzschild solution is recovered. For the very large scale limit, the solution becomes the Schwarzschild-dS solution for $1+\alpha+\beta<0$ and  becomes the Schwarzschild-AdS solution for $1+\alpha+\beta>0$. Moreover, the global monopole solution can be obtained by setting $1+2\alpha+3\beta=0$. Note that the last term in Eq.~\eqref{f sol}, the constant potential $\zeta$, corresponds to the global monopole term which naturally emerges from the graviton mass. Finally, the linear term $\gamma r$ is a characteristic term of this solution, which distinguishes it from other solutions found in literature.

\begin{figure}[h!]
\begin{center}
\includegraphics[scale=0.5]{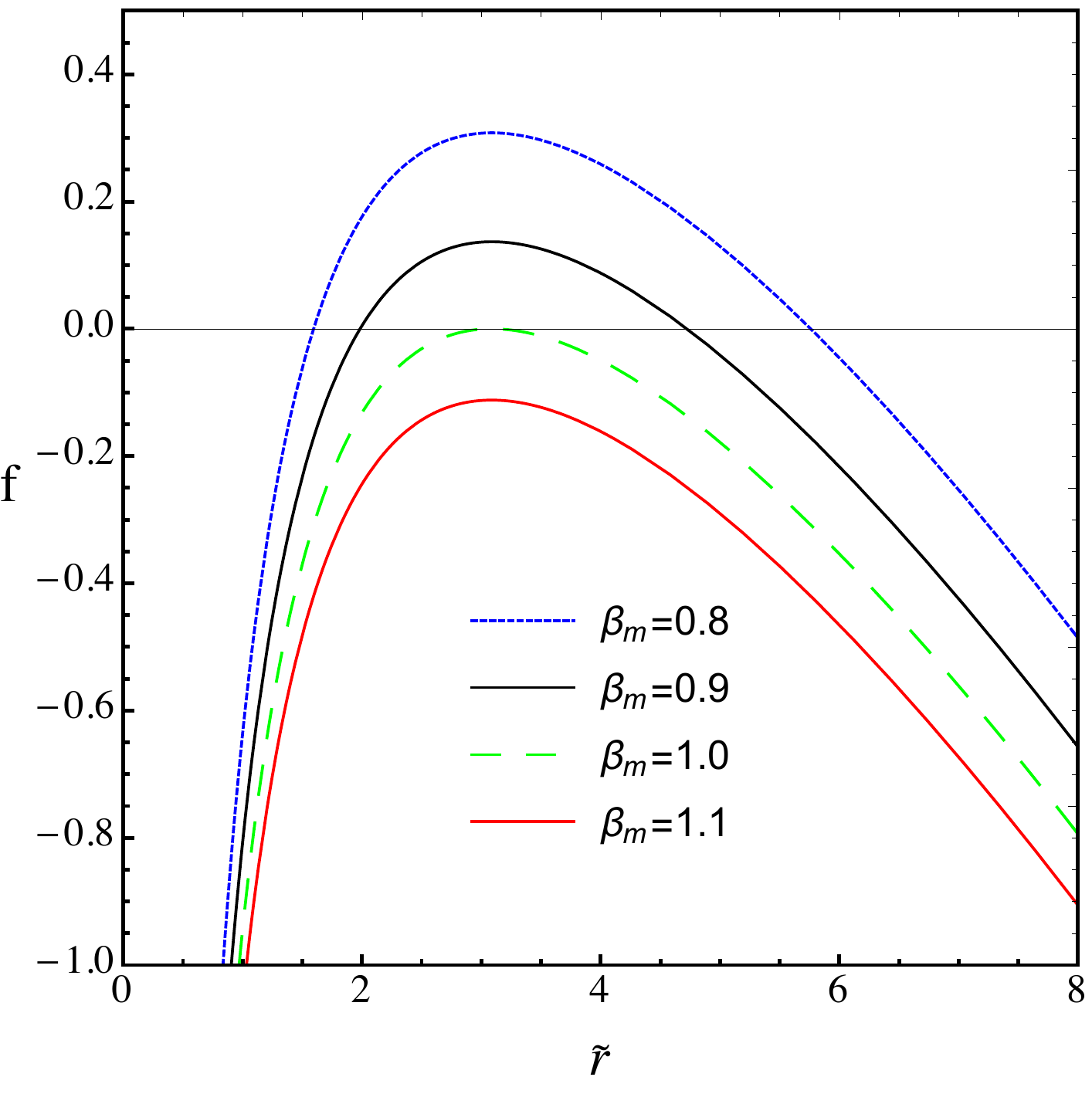}\qquad
\includegraphics[scale=0.5]{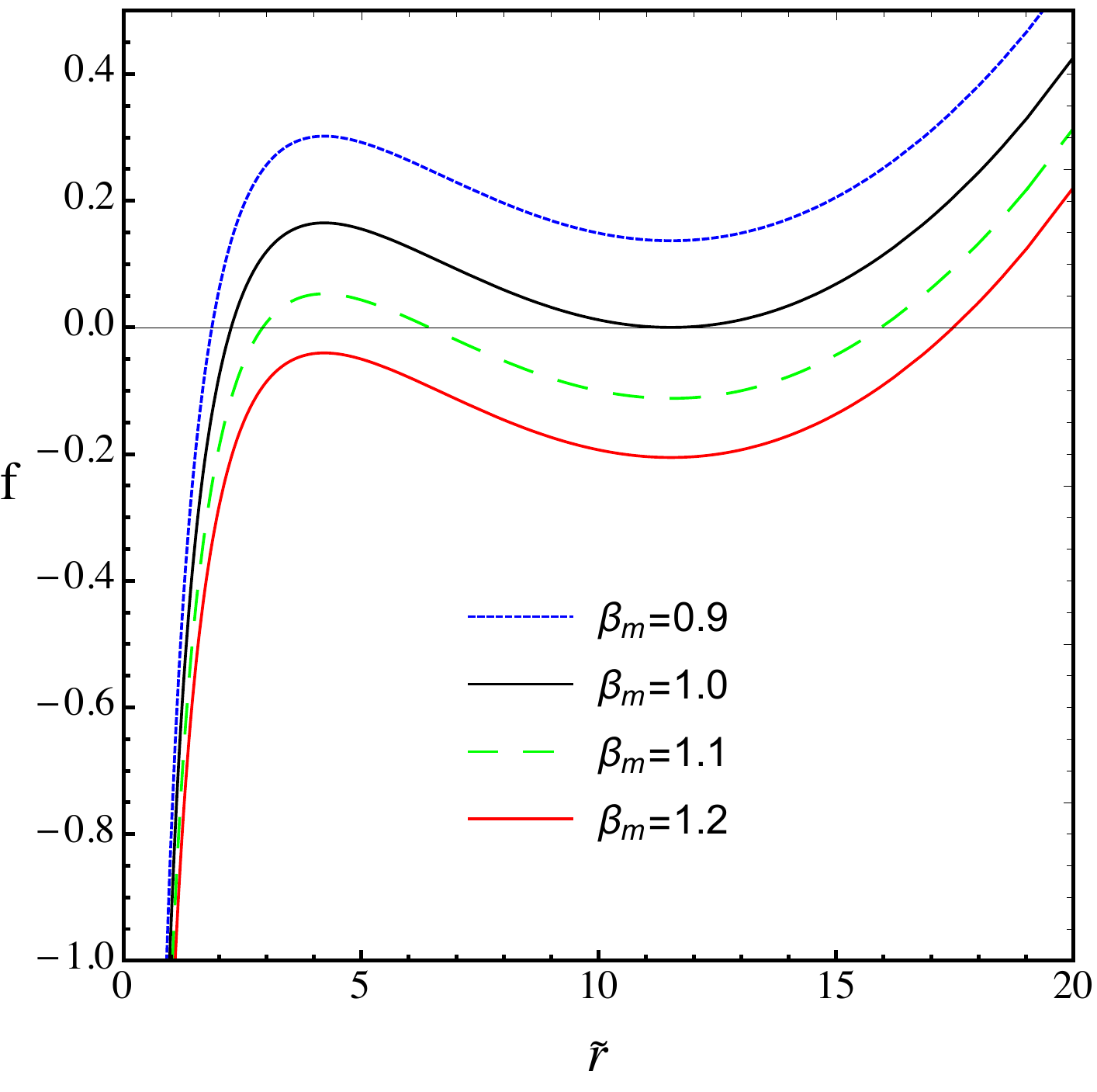}
\end{center}
{\caption{The left-(right-)hand panel shows the horizon structure of the asymptotically dS(AdS) solution in the dRGT massive gravity for various values of $\beta_m$. 
We set the dimensionless parameters as $\tilde{M}=1$, $\alpha_g = 1$ and $c_2 = -0.02/3 (+0.02/3)$ for the asymptotically dS(AdS) solution.
 }\label{fig:Horizon}}
\end{figure}

It is convenient to introduce the dimensionless variable $\tilde{r} = r/h$ and to introduce the dimensionless model parameters \cite{Boonserm:2017qcq}
\begin{eqnarray}
	\tilde{M}&=& \frac{M}{h},\,\,\,\,\,\,
	\alpha_g = m^2_gh^2,\,\,\,\,\,\,
	c_0 = \alpha+3\beta,\,\,\,\,\,\,
	c_1=1 +2\alpha +3\beta,\,\,\,\,\,\,
	c_2 = 1+\alpha+\beta.
\end{eqnarray}
As a result, the function $f$ can be written in terms of the dimensionless variables as
\begin{equation}
	f(\tilde{r}) = 1 - \frac{2\tilde{M}}{\tilde{r}} + \alpha_g \left(c_2\tilde{r}^2 -c_1 \tilde{r}+c_0\right).\label{fdRGT}
\end{equation}
From this equation, one finds that the parameter $h$ characterizes the nonlinear scale of the solution and takes place at $\tilde{M} \sim \alpha_g$. Hence, one can consider the parameter $h$ as
\begin{eqnarray}
	h = r_V = \left(\frac{M}{m_g^2}\right)^{1/3}.
\end{eqnarray}
This radius is well known as the Vainshtein radius \cite{Vainshtein:1972sx}. In the range $r<r_V$, the theory approaches GR, while in the range $r > r_V$, the modification of GR will be active. In order to see the structure of the black hole horizon clearly, let us consider a subclass of parameters by specifying the parameter as follows
\begin{eqnarray}\label{c1c0}
	c_1 = 3 (4c^2_2)^{1/3},\quad c_0 = \frac{9}{\sqrt{3}} \frac{\left(2|c_2|\right)^{1/3}}{\beta_m} - \frac{1}{\alpha_g}.
\end{eqnarray}
Now we have only two significant parameters: $c_2$ and $\beta_m$ characterizing the strength of the graviton mass and the numbers of the horizons, respectively. For asymptotically dS solution, $0 < \beta_m < 1$ is the condition for existence of two horizons while the asymptotically AdS solution, $1 < \beta_m < 2/\sqrt{3}$ is the condition for existence of three horizons. This behavior can be found explicitly using numerical methods as illustrated in Fig \ref{fig:Horizon}.

Moreover, this subclass of parameters allows us to find the exact solutions of the horizon as follows
\begin{footnotesize}
\begin{eqnarray}
	\tilde{r}_{1(dS)} &=& \frac{2}{\left(-2c_2\right)^{1/3}}\left[\left(\frac{2\sqrt{3}}{\beta_m} + 4\right)^{1/2}\cos\left(\frac{1}{3}\sec^{-1}\left(-\frac{\sqrt{\frac{\sqrt{3}}{\beta_m} + 2}\left(2\sqrt{2}\beta_m + \sqrt{6}\right)}{5\beta_m + 3\sqrt{3}}\right)\right) - 1\right],\label{rHdS1}\\
	\tilde{r}_{2(dS)} &=& \frac{-2}{\left(-2c_2\right)^{1/3}}\left[\left(\frac{2\sqrt{3}}{\beta_m} + 4\right)^{1/2}\cos\left(\frac{1}{3}\sec^{-1}\left(-\frac{\sqrt{\frac{\sqrt{3}}{\beta_m} + 2}\left(2\sqrt{2}\beta_m + \sqrt{6}\right)}{5\beta_m + 3\sqrt{3}}\right) + \frac{\pi}{3}\right) + 1\right],\\\label{rHdS2}
	\tilde{r}_{1(AdS)} &=& \frac{2}{\left(2c_2\right)^{1/3}}\left[1 - \left(\frac{4 - 2\sqrt{3}}{\beta_m}\right)^{1/2}\sin\left(\frac{1}{3}\sec ^{-1}\left(\frac{\sqrt{6}-2 \sqrt{2} \beta _m}{\left(3 \sqrt{3}-5 \beta _m\right) \sqrt{-\frac{\beta _m}{\sqrt{3}-2 \beta _m}}}\right) + \frac{\pi}{6}\right)\right],\label{rHAdS1}\\
	\tilde{r}_{2(AdS)} &=& \frac{2}{\left(2c_2\right)^{1/3}}\left[1 - \left(\frac{4 - 2\sqrt{3}}{\beta_m}\right)^{1/2}\cos\left(\frac{1}{3}\sec^{-1}\left(\frac{\sqrt{6} - 2\sqrt{2}\beta _m}{\left(3 \sqrt{3}-5 \beta _m\right) \sqrt{-\frac{\beta _m}{\sqrt{3}-2 \beta _m}}}\right) + \frac{\pi}{3}\right)\right],\label{rHAdS2}\\
	\tilde{r}_{3(AdS)} &=& \frac{2}{\left(2c_2\right)^{1/3}}\left[1 + \left(\frac{4 - 2\sqrt{3}}{\beta_m}\right)^{1/2}\cos\left(\frac{1}{3}\sec^{-1}\left(\frac{\sqrt{6} - 2\sqrt{2}\beta_m}{\left(3\sqrt{3} - 5\beta_m\right)\sqrt{-\frac{\beta_m}{\sqrt{3} - 2\beta_m}}}\right)\right)\right].\label{rHAdS3}
\end{eqnarray}
\end{footnotesize}


\section{Dirac perturbation}\label{sec:eff potential}


In this section, the effective potential of QNM is derived in the spherically symmetric spacetime for arbitrary gravity theories. It is then expressed, exactly, for the dRGT massive gravity, using the solution (\ref{fdRGT}). Let us start with the general form of the metric
\begin{eqnarray}
	\text{d}s^2=-f(r)\text{d}t^2+\frac{1}{f(r)}\text{d}r^2+r^2(\text{d}\theta^2+\sin^2\theta\text{d}\phi^2).
\end{eqnarray}
For spin-half fields in curved spacetime, it is convenient to consider the calculation in the vielbein formalism. In this present work, we choose the form of the vielbein as follows
\begin{eqnarray}
	e^\mu_{\,\,\,\hat{\alpha}}&=&\text{diag}\left(\frac{1}{\sqrt{f}}, \sqrt{f}, \frac{1}{r}, \frac{1}{r\sin\theta}\right).
\end{eqnarray}
Note that the indices without hats are curved spacetime indices and ones with hats are Lorentz indices. We consider a test spin-1/2 field near the spherically symmetric spacetime as a perturbed field. Therefore, one can fix the background and then the backreaction can be neglected. The Dirac equation in general curved spacetime is expressed as
\begin{eqnarray}
	\Big[\gamma^\mu(\partial_\mu+\Gamma_\mu)+m\Big]\Psi=0,\label{Dirac eq}
\end{eqnarray}
where $\Psi$ and $m$ are the Dirac field and its mass respectively. $\gamma^\mu$ is the $4\times4$ Dirac gamma matrix and $\Gamma_\mu$ is the spin connection given by
\begin{eqnarray}
	\Gamma_\mu&=&\frac{1}{2}\,\omega_{\mu\hat{\alpha}\hat{\beta}}\,\Sigma^{\hat{\alpha}\hat{\beta}},
\end{eqnarray}
with
\begin{eqnarray}
	\omega_{\mu\hat{\alpha}\hat{\beta}}=e^\rho_{\,\,\,\hat{\alpha}}(\partial_\mu e_{\rho\hat{\beta}}-\Gamma^\sigma_{\mu\rho}e_{\sigma\hat{\beta}}),\hspace{1cm}
	\Sigma^{\hat{\alpha}\hat{\beta}}=\frac{1}{4}[\gamma^{\hat{\alpha}},\gamma^{\hat{\beta}}].
\end{eqnarray}
$\Gamma^\rho_{\mu\nu}$ is the Christoffel symbol. The representation of the Dirac gamma matrices, $\gamma^{\hat{\alpha}}$ is chosen as follows \cite{Hung2015}
\begin{eqnarray}
	\gamma^{\hat{0}}=i\sigma^3\otimes\mathbbm{1},\hspace{1cm}
	\gamma^{\hat{1}}=\sigma^2\otimes\mathbbm{1},\hspace{1cm}
	\gamma^{\hat{2}}=\sigma^1\otimes\sigma^1,\hspace{1cm}
	\gamma^{\hat{3}}=-\sigma^1\otimes\sigma^2,
\end{eqnarray}
with the Pauli spin matrices:
\begin{eqnarray}
	\sigma^1=\left(\begin{array}{cc}0&1\\1&0\end{array}\right),\hspace{1cm}
	\sigma^2=\left(\begin{array}{cc}0&-i\\i&0\end{array}\right),\hspace{1cm}
	\sigma^3=\left(\begin{array}{cc}1&0\\0&-1\end{array}\right).
\end{eqnarray}
Using these, Eq. \eqref{Dirac eq} can be reexpressed as
\begin{eqnarray}
	\left[
	\frac{1}{\sqrt{f}}(i\sigma^3\otimes\mathbbm{1})\partial_t
	+\sqrt{f}(\sigma^2\otimes\mathbbm{1})\partial_r
	+\frac{1}{r}(\sigma^1\otimes\sigma^1)\partial_\theta
	-\frac{1}{r\sin\theta}(\sigma^1\otimes\sigma^2)\partial_\phi
	\right.\hspace{.6cm}&&\nonumber\\
	\left.
	+\frac{f'}{4\sqrt{f}}(\sigma^2\otimes\mathbbm{1})
	+\frac{\sqrt{f}}{r}(\sigma^2\otimes\mathbbm{1})
	+\frac{\cot\theta}{2r}(\sigma^1\otimes\sigma^1)	
	+m\right]\Psi&=&0.\,\,\,
\end{eqnarray}
where the prime denotes the derivative with respect to $r$.

One of the ways to solve this equation is using the separation method. Since the metric admits spherical symmetry and does not depend on $t$, it is possible to separate the solution into the radial, temporal and angular parts such that
\begin{eqnarray}
	\Psi(t,r,\theta,\phi)=\left(\begin{array}{c}iA(r)\\B(r)\end{array}\right)e^{-i\omega t}\otimes\Theta(\theta,\phi),
\end{eqnarray}
where $A$ and $B$ are the radial functions, and $\omega$ is the angular frequency of the solution. The spherically symmetric angular part $\Theta$ satisfies the eigen equation for the Dirac field on a two-dimensional sphere,
\begin{eqnarray}
	\left(\sigma^1\partial_\theta-\frac{\sigma^2}{\sin\theta}\partial_\phi+\frac{\cot\theta}{2}\sigma^1\right)\Theta=i\lambda\Theta,
\end{eqnarray}
where $\lambda=\pm1,\pm2,\pm3,\hdots$ are the corresponding eigenvalues. As a result, the radial equation for $A$ and $B$ can be written as
\begin{eqnarray}
	\left[\left(f\partial_r+\frac{f'}{4}+\frac{f}{r}\right)\sigma^2
	+\frac{i\lambda\sqrt{f}}{r}\sigma^1\right]
	\left(\begin{array}{c}iA\\B\end{array}\right)
	&=&
	-\left[\omega\sigma^3+m\sqrt{f}\mathbbm{1}\right]
	\left(\begin{array}{c}iA\\B\end{array}\right).\label{Dirac eq2}
\end{eqnarray}
This radial equation is still complicated. In order to simplify it, one can introduce a function $C(r)$ to eliminate the terms $\frac{f'}{r}+\frac{f}{r^2}$. This function must satisfy the condition, $f\partial_rC+\frac{f'}{4}C+\frac{f}{r}C=0$. For the case of the dRGT massive gravity, $C(r)$ takes the form
\begin{eqnarray}
	C(r)=C_0r^{-3/4}\Big[r \left(\Lambda r^2+3\gamma r+3\zeta+3\right)-6 M\Big]^{-1/4},
\end{eqnarray}
where $C_0$ is an integration constant. By setting
\begin{eqnarray}
	\left(\begin{array}{c}B/C\\A/C\end{array}\right)
	=\left(\sin\frac{\theta}{2}\,\sigma^3+\cos\frac{\theta}{2}\,\sigma^1\right)\left(\begin{array}{c}\tilde{B}\\\tilde{A}\end{array}\right),
\end{eqnarray}
where $\theta=\tan^{-1}(-mr/\lambda)$, Eq. \eqref{Dirac eq2} is then simplified as
\begin{eqnarray}
	f\partial_r\tilde{B}+a\tilde{B}&=&-\omega b\tilde{A},\\
	f\partial_r\tilde{A}-a\tilde{A}&=&\omega b\tilde{B},
\end{eqnarray}
where
\begin{eqnarray}
	a=\frac{\sqrt{f}}{r}\sqrt{\lambda^2+m^2r^2},
	\hspace{1cm}
	b=1+\frac{fm\lambda}{2\omega(m^2r^2+\lambda^2)}.
\end{eqnarray}
Let us introduce the new coordinate $x$ called tortoise coordinate. This coordinate is related to the radial coordinate $r$ via
\begin{eqnarray}
	f\,\partial_r=b\,\partial_{x}.
\end{eqnarray}
Note that $x\to-\infty$ as $r$ goes to the event horizon and $x\to\infty$ as $r$ goes to the cosmological horizon. The range of the new coordinate is thus expanded to be from $-\infty$ to $\infty$. Eventually, the decoupled radial equations are obtained as
\begin{eqnarray}
	\left(-\partial^2_{x}+V_+\right)\tilde{A}&=&\omega^2\tilde{A},\\
	\left(-\partial^2_{x}+V_-\right)\tilde{B}&=&\omega^2\tilde{B}.
\end{eqnarray}
Note that these equations are Schr\"{o}dinger-like equations with effective potentials,
\begin{eqnarray}
	V_\pm=\pm\partial_{x}\left(\frac{a}{b}\right)+\left(\frac{a}{b}\right)^2.
\end{eqnarray}
We also note that although, there are two potentials, $V_+$ and $V_-$, obtained from the same function $a/b$, called the superpotential. This means that the potentials $V_+$ and $V_-$ are supersymmetric partners \cite{CKS1995}. They thus give the same spectra of QMNs \cite{Zhou2014}.

For the case of the massless Dirac field ($m=0$), the potential is expressed as
\begin{eqnarray}\label{Veff}
	V_\pm
	&=&\pm f\,\partial_r\left(\frac{\sqrt{f}}{r}\lambda\right)+\frac{f}{r^2}\lambda^2.
\end{eqnarray}
Substituting $f=1-\frac{2 M}{r}+\frac{\Lambda}{3}r^2+\gamma r+\zeta$, the potential can be written as
\begin{eqnarray}
	V_\pm
	&=&\pm f\left[\lambda\,\left(\frac{6\mu-r(2+\gamma r+2\zeta)}{2r^3\sqrt{f}}\right)+\lambda^2\left(\frac{1}{r^2}\right)\right]. \label{potential}
\end{eqnarray}
Note that Eq.~(\ref{Veff}) is consistent with the Dirac perturbation in GR cases when the dRGT parameters vanish \cite{Cho:2005yc}. In this work, one chooses to study a QMN with the potential $V_+$. By using the parameters defined in the previous section, one finds that there are three crucial parameters, $c_2$, $\beta_m$ and $\lambda$. Moreover, we will see that the potential vanishes at the horizon since $f=0$, and approaches a constant value in the large $r$ (or $\tilde{r}$) limit. Note that the potential is valid only in the range $f\geq0$ due to the existence of $\sqrt{f}$.  

\begin{figure}
\begin{center}
\includegraphics[scale=0.48]{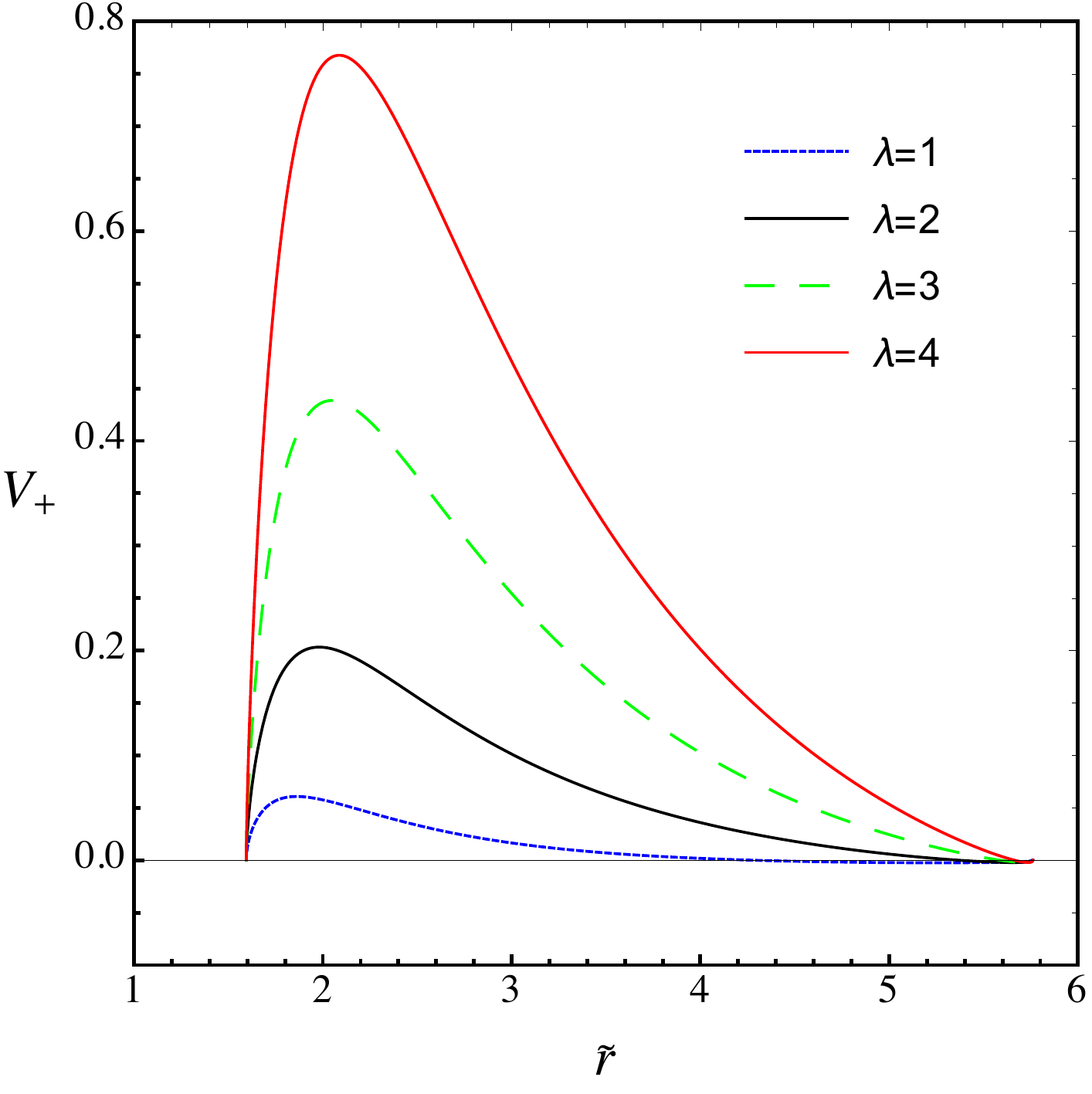}\qquad
\includegraphics[scale=0.5]{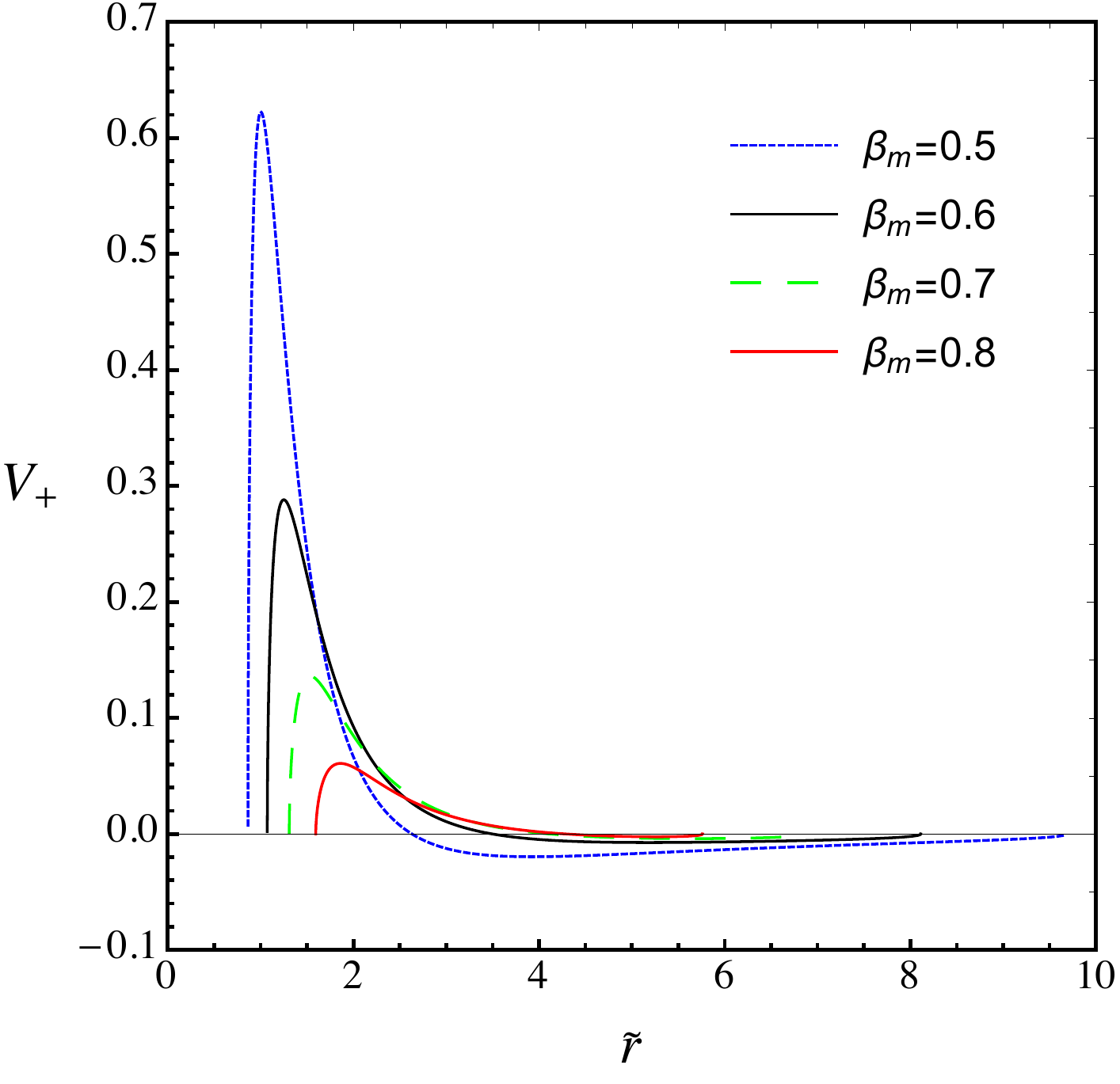}
\end{center}
{\caption{The left-hand panel shows the potential for different values of the parameter $\lambda$,\\ with $\beta_m =0.8$ and $c_2 =-0.02/3 $. The right-hand panel shows the potential \\for different values of $\beta_m$ with $\lambda =1$ and $c_2 =-0.02/3 $.}\label{fig:V-lam-bm}}
\end{figure}

By Eq.~\eqref{potential}, it is obvious that the potential is higher when the parameter $\lambda$ is larger. This behavior is also illustrated in the left-hand panel of Fig. \ref{fig:V-lam-bm}. From the right-hand panel of this figure, one can see that the potential becomes lower when the parameter $\beta_m$ increases closer to $1$.

\begin{figure}[h!]
\begin{center}
\includegraphics[scale=0.5]{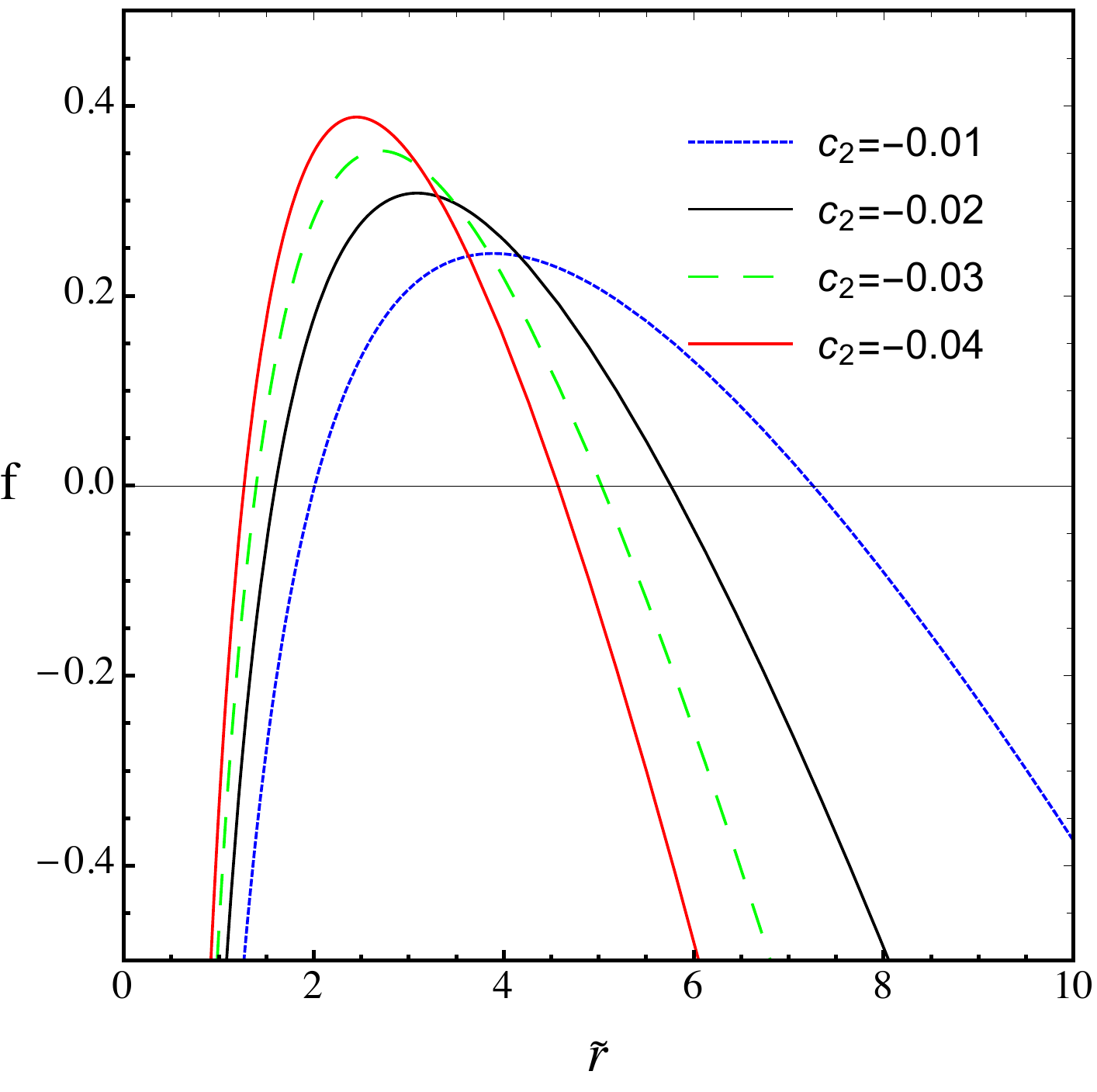}\qquad
\includegraphics[scale=0.5]{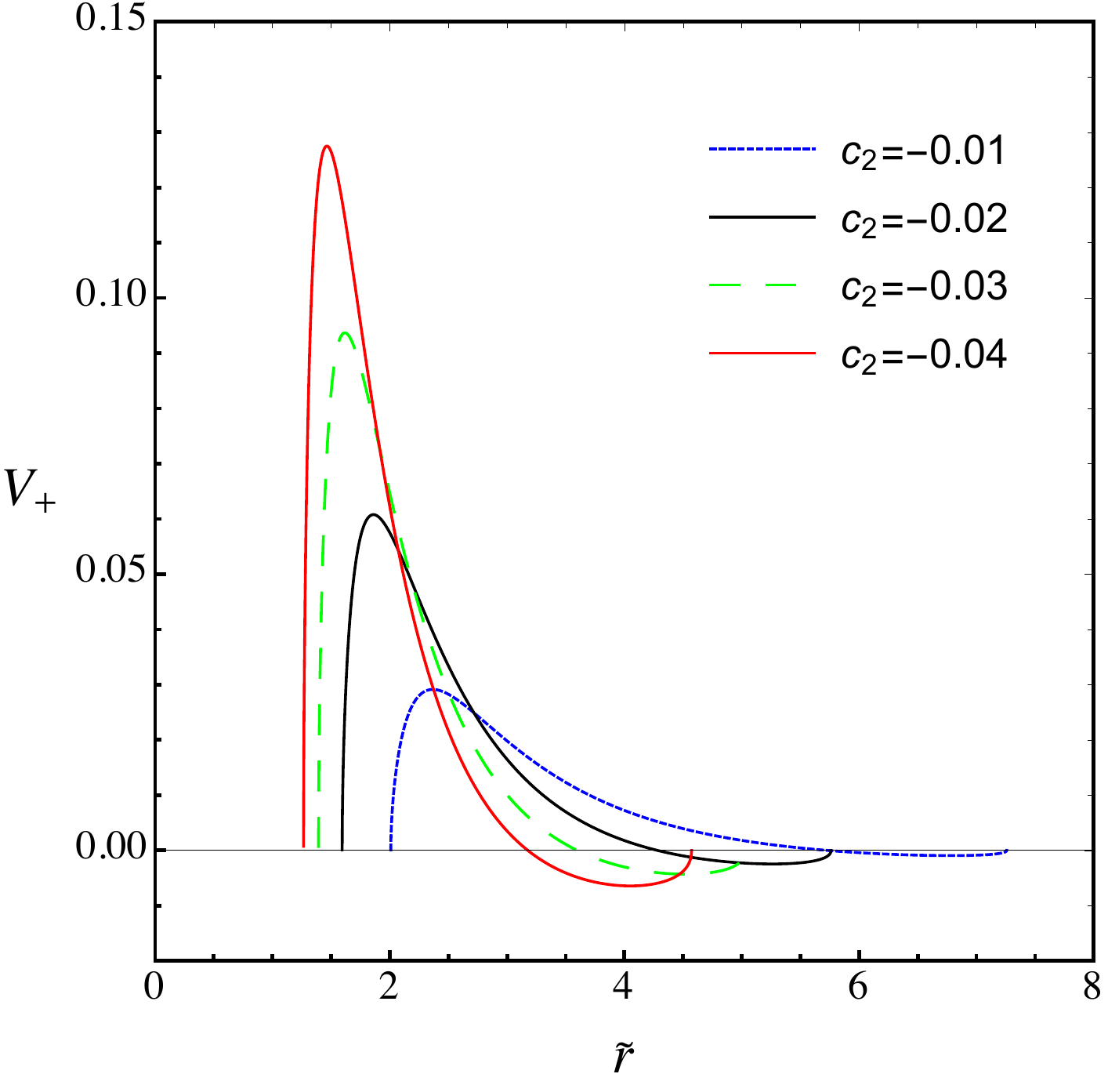}
\end{center}
{\caption{The left-hand panel shows the horizon structure for different values of the parameter $c_2$ with $\beta_m =0.8$ and $\lambda =1$. The right-hand panel shows the potential\\ for different values of the parameter $c_2$ with $\beta_m =0.8$ and $\lambda =1 $.}\label{fig:fV-c2}}
\end{figure}

The parameter $c_2$ controls the strength of the graviton mass or the cosmological constant. As shown in the left-hand panel of Fig. \ref{fig:fV-c2}, smaller values of $|c_2|$ yield larger distance for the cosmological horizon. This behavior is similar to one in cosmological aspect; the cosmological constant is very small and then the cosmological horizon is very far. This means that gravity is significantly modified only on very large scales. As a result, the potential becomes wider and lower when $|c_2| \rightarrow 0$, as shown in the right-hand panel of Fig. \ref{fig:fV-c2}.

\section{Quasinormal modes}\label{sec:QNM}
In the black hole perturbation theory with spherically symmetric spacetime, the radial equations always can be represented in a Schr\"{o}dinger-like form with an effective potential. The effective potential is determined by the specific spin of the particle and the specific kind of black hole. The QNM spectrum is one of the important physical properties of the wave which can be obtained by solving the radial equation. Therefore, the shape of the potential is important to characterize the QNM and vice versa.

In a rough classification, there are two main types of effective potential corresponding to various fields in spherically symmetric spacetimes. The first is the barrier-like effective potential, which includes a local maximum and is asymptotically zero, or converges to a specific value which is smaller than the maximum at spatial infinity or the cosmological horizon. The QNMs can be obtained by taking the boundary conditions of purely ingoing waves (i.e. ingoing to the black hole event horizon) and purely outgoing waves (i.e. outgoing to spatial infinity or to the cosmological horizon). The physical phenomena of QNMs associated with this type of potential represent the ringdown profile of black hole merges and the damping rate of the late-time tail of the propagating waves (i.e. waves with positive real and negative imaginary frequencies). The second type of effective potential is asymptotic infinity or a finite maximum at spatial infinity, which usually relates to the AdS black hole. The boundary conditions for obtaining the QNMs are slightly different from the previous type of potential. In this case, we require the purely ingoing waves to the black hole event horizon, and Dirichlet or vanishing energy flux boundary conditions at spatial infinity. The QNMs of this potential are suggested to link with the AdS/CFT correspondence. For a more detailed discussion of the black hole QNMs, we refer to these review articles \cite{BCS2009,KZ2011}.

The dRGT black hole model contains both types of effective potential. The asymptotically AdS case contains, however, more complicated structures \cite{Ghosh:2015cva}. The study of QNMs for the asymptotically AdS case is model dependent and beyond the scope of our current study. As a result, we focus on the dS-like (asymptotically dS) solutions in which the effective potential is always barrier-like, as mentioned in the previous section. In this case, it is useful to study how the structure of the graviton mass can provide significant deviations from the Schwarzschild-dS case.

We study the Dirac QNMs using the 3rd order WKB approximation by Iyer and Will \cite{iyewil}, the 6th order WKB approximation by Konoplya \cite{kono2003}, and the recent revised WKB approach with Pad$\acute{e}$ approximation by Konoplya, Zhidenko and Zinhailo \cite{KZZ2019}. These methods are powerful tools in the study of barrier-like potentials with the aforementioned boundary conditions. We choose the effective potential $V_{+}$ in Eq.~(\ref{Veff}) to evaluate the QNMs while the potential $V_{-}$ will be the super-symmetric partner potential as mentioned in the previous section. The metric element $f$ is given by Eq.~(\ref{fdRGT}). We take the black hole mass $\tilde{M}=1$, the scaling parameter $\alpha_{g}=1$, and set the parameters $c_{0}$ and $c_{1}$ as presented in Eq.~(\ref{c1c0}). There are three leftover parameters containing crucial physical meanings as follows: the angular momentum parameter $\lambda=l+1$, $l=0,\ \pm1,\ \pm2,\hdots$ which is based on the spin-1/2 eigenvalues on the two-dimensional sphere, the strength parameter of the graviton mass, $c_{2}$, which is analogous to the cosmological constant in traditional GR, and the free parameter $\beta_{m}$ which allows the effective potential to always be dominated by barrier-like behavior between the black hole and cosmological horizons, when $0<\beta_{m}<1$.

\begin{figure}[h!]
\begin{center}
\includegraphics[scale=0.6]{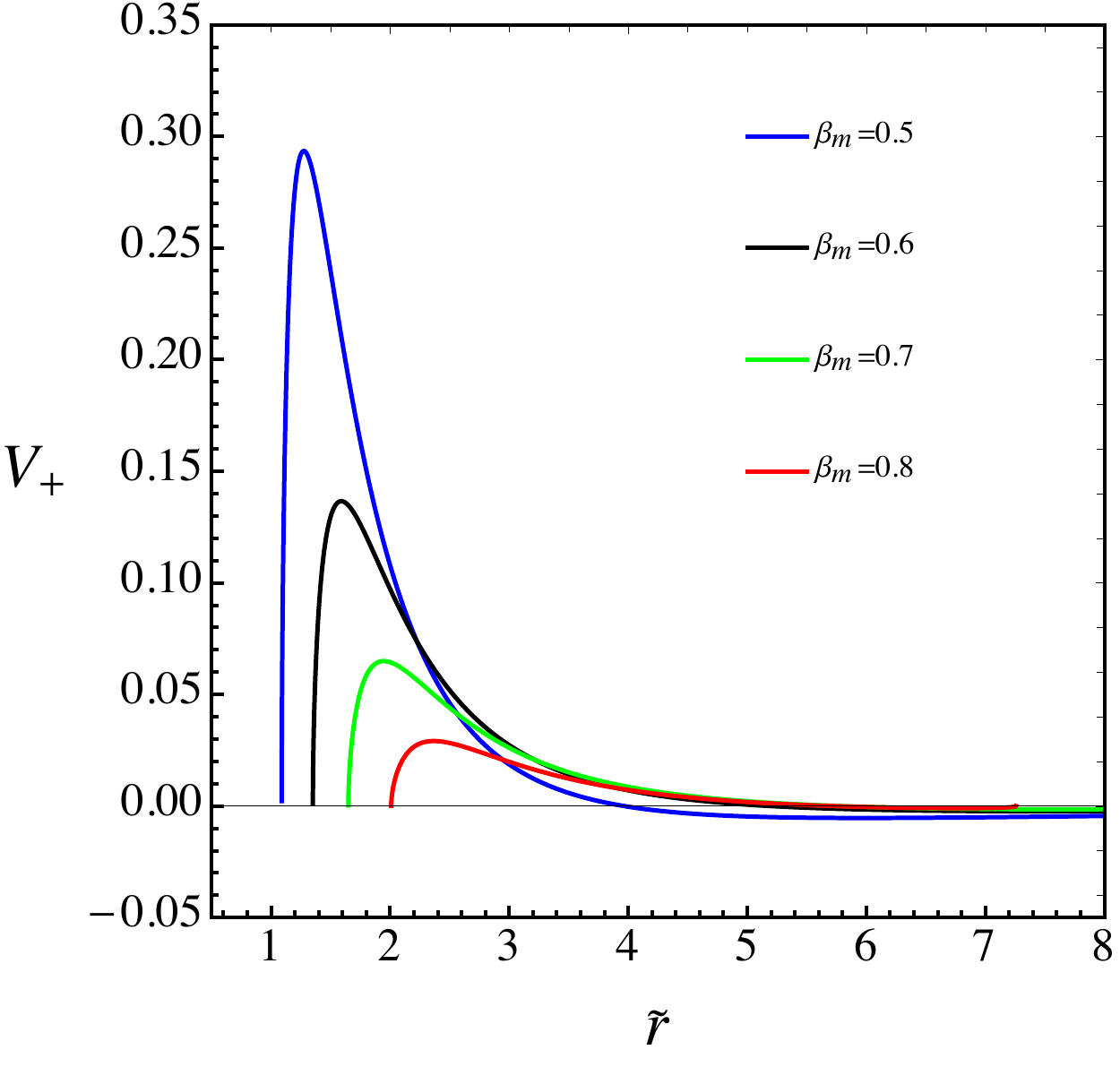}\qquad
\includegraphics[scale=0.8]{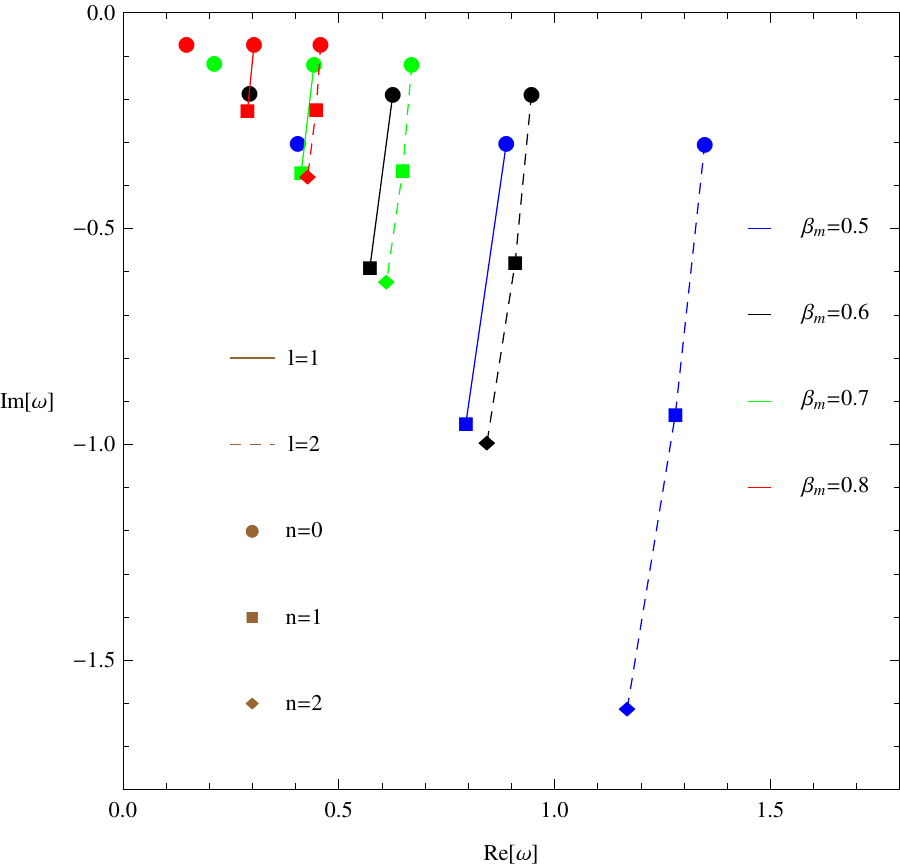}
\end{center}
{\caption{The left-hand panel shows the effective potential with $c_{2}=-\frac{0.01}{3}$, $l=0$ and \\various low values of $\beta_{m}$. The right-hand panel shows the related low-lying QNMs.}\label{fig:QNM1}}
\end{figure}

\begin{figure}[h!]
\begin{center}
\includegraphics[scale=0.6]{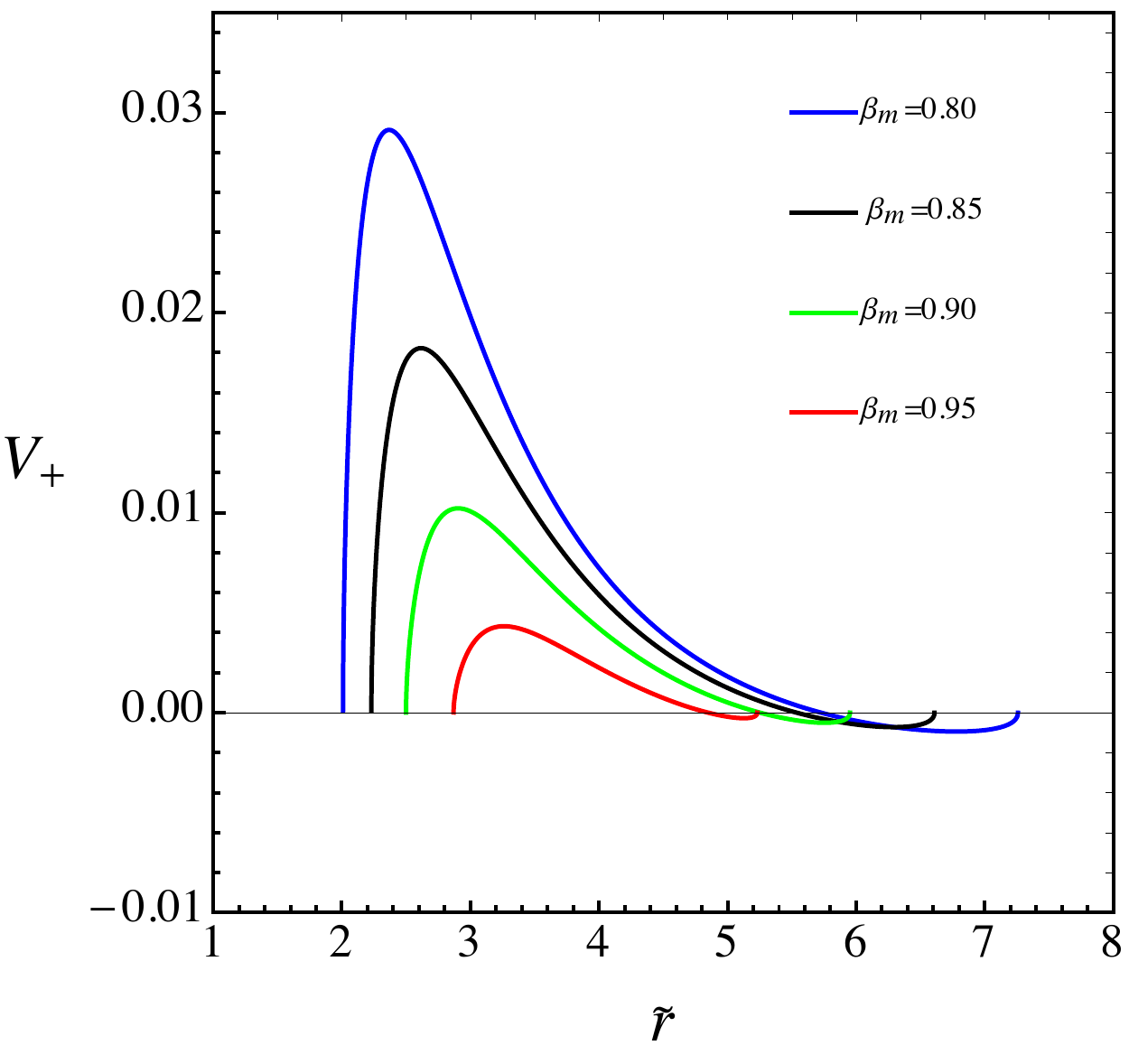}\qquad
\includegraphics[scale=0.8]{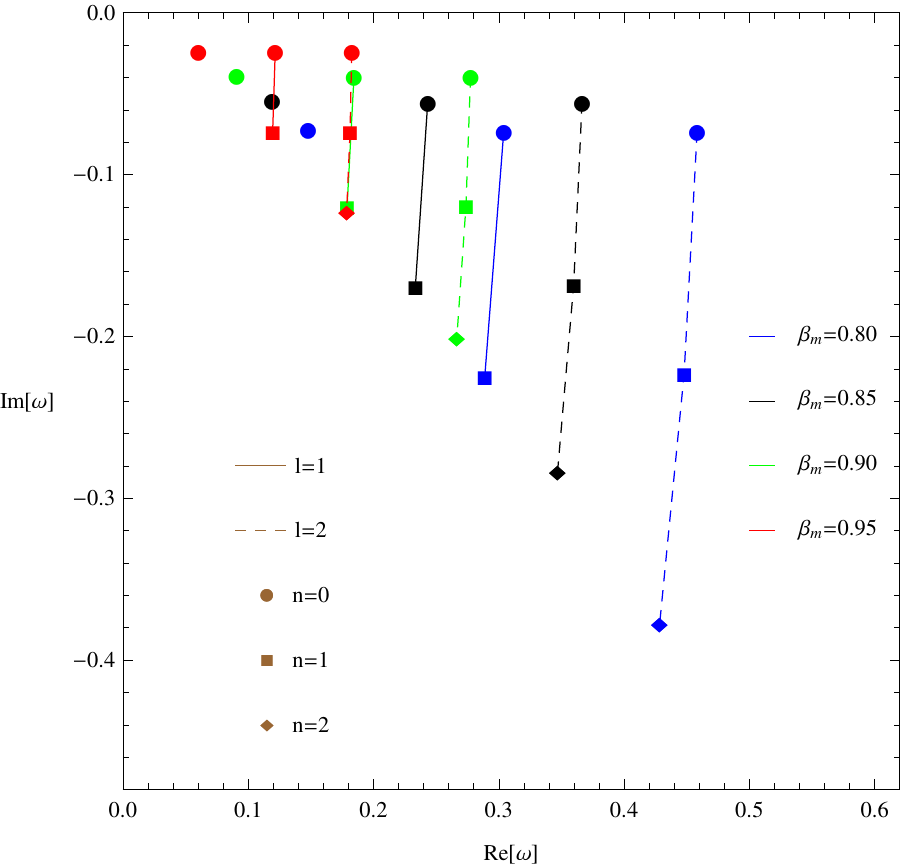}
\end{center}
{\caption{The left-hand panel shows the effective potential with $c_{2}=-\frac{0.01}{3}$, $l=0$ and \\various high values of $\beta_{m}$. The right-hand panel shows the related low-lying QNMs.}\label{fig:QNM2}}
\end{figure}

\begin{figure}[h!]
\begin{center}
\includegraphics[scale=1]{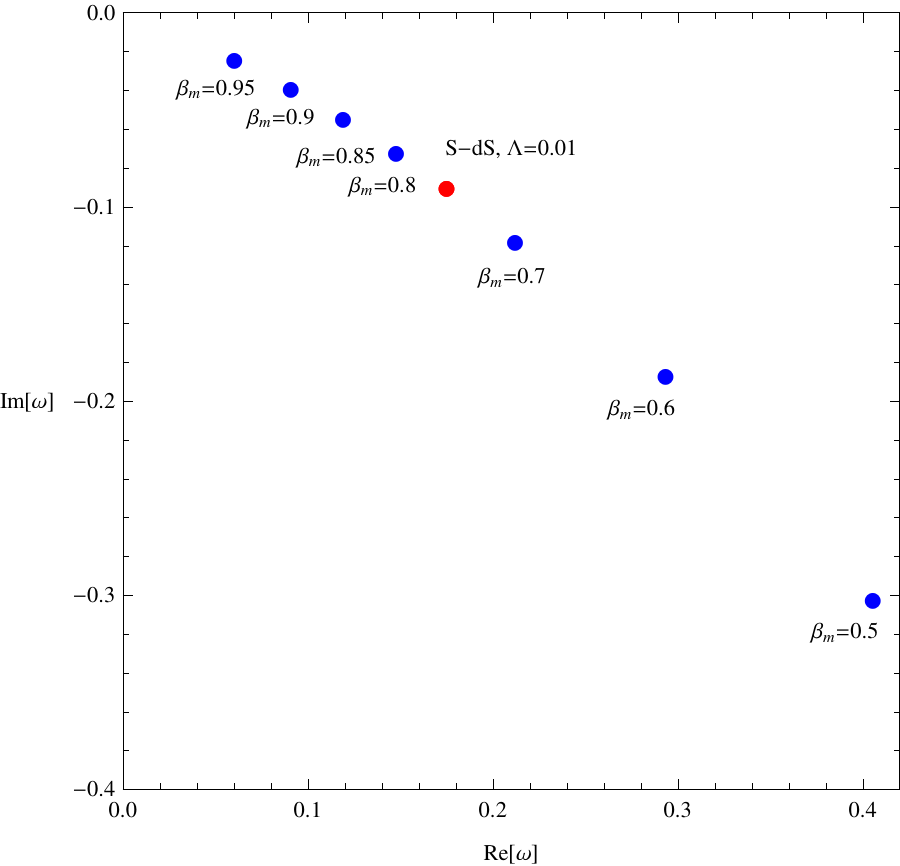}
\end{center}
{\caption{Comparison of the lowest QNM frequency ($n=0$ and $l=0$) for the dRGT \\ black hole with $c_{2}=-\frac{0.01}{3}$ and one for the Schwarzschild de-Sitter solution, with $\Lambda=0.01$.}\label{fig:QNM3}}
\end{figure}

We evaluate two different sets of parameters for the QNMs. In the first set, we fix $c_{2}=-\frac{0.01}{3}$ and vary the parameter $\beta_m$ as follows: $\beta_{m}=\ 0.5,\ 0.6,\ 0.7.\ 0.8,\ 0.85, \ 0.9$ and $0.95$. This set of parameters allows us to compare, straightforwardly, the results in the dRGT case with these for the Schwarzschild-dS solution when the cosmological constant is $\Lambda=0.01$. The low-lying modes with the 3rd and 6th order WKB approximation, the 6th and 13th order revised WKB approaches with Pad$\acute{e}$ approximation, as well as the reference modes of the Schwarzschild-dS cases are explicitly presented in Tables.~\ref{Tab1}, \ref{Tab2}, \ref{Tab3}, \ref{Tab4}, \ref{Tab5} and \ref{Tab6} in Appendix A. The relations for the corresponding effective potential and the evolution of these modes with the 6th order WKB approximation are also presented in Figs.~\ref{fig:QNM1} and \ref{fig:QNM2}, for low and high $\beta_m$, respectively. For fixed $l$ and $n$, where $n$ is the mode number, the QNM frequencies shift to the smaller real part and smaller absolute values of the imaginary part when $\beta_{m}$ increases. This means that the propagating wave of QNM will oscillate and decay slower for lower effective potentials. Moreover, it is found that, for fixed $l$ and increasing $n$, the real parts of the QNM frequencies decrease while the absolute values of the imaginary parts increase. This implies that the oscillating frequency of the propagating wave is smaller and the damping rate becomes larger for higher $n$ modes. These properties can be checked analytically for the leading 3rd order term of the WKB approximation with the form of the frequency as follows,
\begin{equation}
	\omega^{2}=\left[V_{0}+\left(-2V''_{0}\right)^{1/2}\Lambda\right]-i\left(n+\frac{1}{2}\right)\left(-V''_{0}\right)^{1/2}\left(1+\Omega\right).\label{3rd WKB freq}
\end{equation}
Here, $V_{0}$ denotes the maximum of $V_{+}$, and the prime now denotes the derivative with respective to the tortoise coordinate. The functions $\Lambda$ and $\Omega$ can be expressed as
\begin{eqnarray}
	\Lambda
	&=&\frac{1}{\left(-2V''_{0}\right)^{1/2}}\left[\frac{1}{8}\left(\frac{V_{0}^{(4)}}{V''_{0}}\right)\left(\frac{1}{4}+\alpha^{2}\right) -\frac{1}{288}\left(\frac{V'''_{0}}{V''_{0}}\right)^{2}\left(7+60\alpha^{2}\right)\right],\nonumber\\
	\Omega
	&=&\frac{1}{\left(-2V''_{0}\right)}\Bigg[\frac{5}{6912}\left(\frac{V'''_{0}}{V''_{0}}\right)^{4}\left(77+188\alpha^{2}\right)-\frac{1}{384} \left(\frac{V_{0}^{'''2} V_{0}^{(4)}}{V_{0}^{''3}}\right)\left(50+100\alpha^{2}\right)\nonumber\\
	&&+\frac{1}{2304}\left(\frac{V_{0}^{(4)}}{V''_{0}}\right)^{2}\left(67+68\alpha^{2}\right)+\frac{1}{288}\left(\frac{V'''_{0}V_{0}^{(5)}}{V_{0}^{'''2}}\right) \left(19+28\alpha^{2}\right)-\frac{1}{288}\left(\frac{V_{0}^{(6)}}{V''_{0}}\right)\left(5+4\alpha^{2}\right)\Bigg],\nonumber\\
\end{eqnarray}
where $\alpha=n+1/2$, $n=0,1,2,...$. One can see that the essential part (leading order) comes from $V_0$ which is the maximum value of the potential. As a result, the imaginary part of $\omega$ is proportional to $V_0$. This makes the waves with QNMs decay faster for higher values of $V_0$. These behaviors are also consistent with the general expectation in traditional GR cases, as in \cite{Zhi2003,Cho:2005yc,CCDW2009}. Comparison of the $n=l=0$ mode with the Schwarzschild-dS case is presented in Fig.~\ref{fig:QNM3}. It is seen that the frequencies obtained from Schwarzschild-dS case are located approximately on the linear region of parameter in the dRGT model. In other word, the results from the Schwarzschild-dS case is a subclass of ones from the dRGT model. For example, the lowest QNM frequency from the Schwarzschild-dS case with $\Lambda=0.01$ corresponds to one from the dRGT case with $c_2=-0.01/3$ and $\beta_m\sim0.77$. This implies that it is possible to obtain faster or slower decay rates of the wave in the dRGT black holes compared to one in the Schwarzschild-dS black hole.


\begin{figure}[h!]
\begin{center}
\includegraphics[scale=0.6]{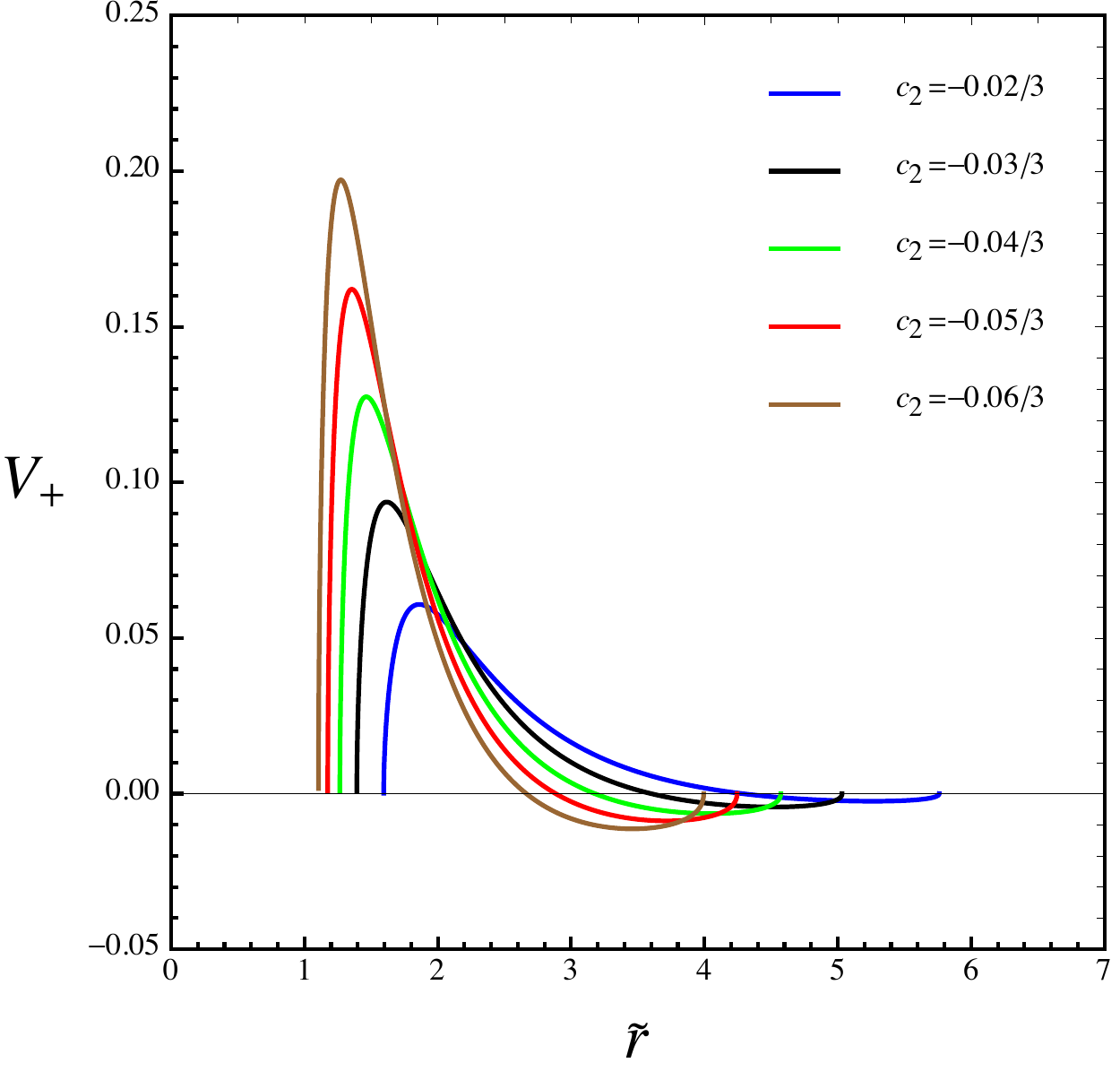}\qquad
\includegraphics[scale=0.8]{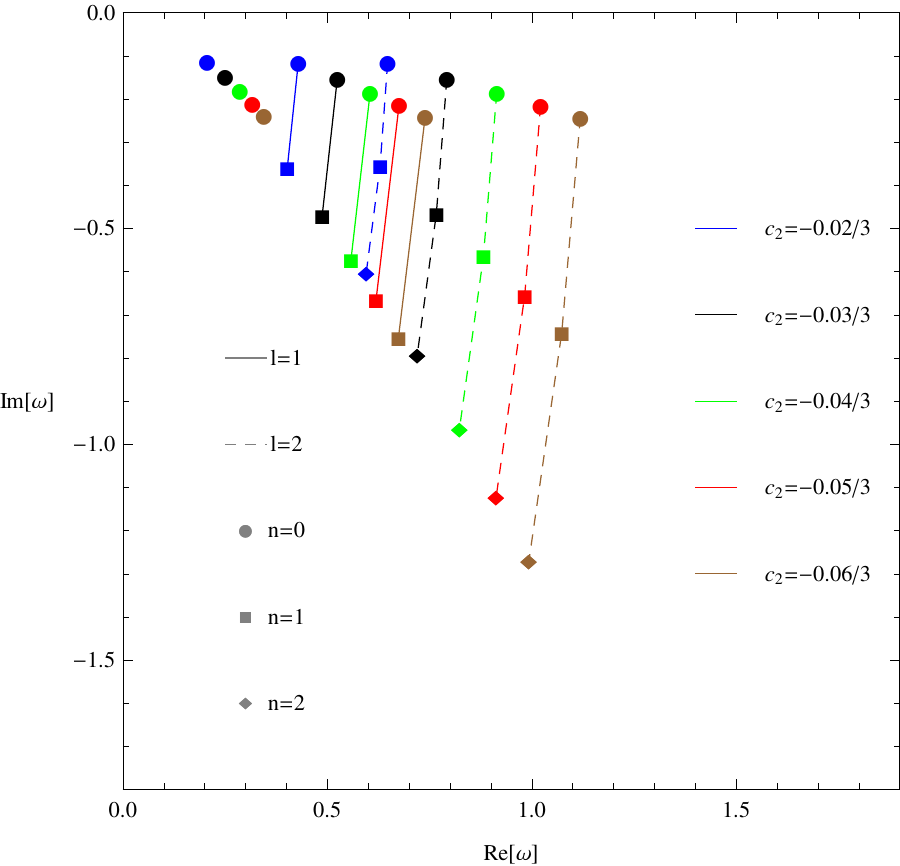}
\end{center}
{\caption{The left-hand panel shows the effective potential with $\beta_{m}=0.8$, $l=0$ and various values of $c_{2}$. The right-hand panel shows the related low-lying QNMs.}\label{fig:QNM4}}
\end{figure}

For the second set, we fix $\beta_{m}$ and vary the parameter $c_2$ as follows $c_{2}=-\frac{0.02}{3},\ -\frac{0.03}{3},\ -\frac{0.04}{3}$, $-\frac{0.05}{3}$ and $-\frac{0.06}{3}$. The results of the low-lying modes with the 3rd and 6th order WKB approximation, and the 6th and 13th order revised WKB approaches with Pad$\acute{e}$ approximation are explicitly listed in Tables.~\ref{Tab7}, \ref{Tab8}, \ref{Tab9}, \ref{Tab10}, \ref{Tab11} and \ref{Tab12} in Appendix A. The 6th order WKB approximation results and the corresponding effective potential are shown in Fig.~\ref{fig:QNM4}. From this figure, one can see that the real parts shift to larger values and the absolute values of imaginary parts become larger for the low-lying modes with fixed $n$ and $l$ and increasing $|c_2|$. Again, the same behavior as the fixed $c_2$ and varying $\beta_m$ case is obtained, the higher potential, the faster of the decay rate of the wave.

It is worthwhile to note that, for the dRGT black hole, there exists a finite potential well between the maximum point and the cosmological horizon. However, the dominant part is still positive then the potential can be approximated as a barrier-like potential. The potential well does not influence the behavior of QNMs evaluated by the WKB formula, as discussed in \cite{KZZ2019}. Note that if the finite potential well is strong enough to influence the QNMs behavior, it will correspond to the “three turning points” effective potential mentioned in \cite{GM1992}. This is worthy of further study as an extension of the present work.

As a remark, the results were evaluated by Wolfram Mathematica with versions $8.0$ to $11.0$. The data is consistent and reproducible only with enough precision numbers of numerical input. This is because the precision setting of Wolfram Mathematica is not sufficient to yield credible results. The details of the results for different precision setting used in numerical calculations are listed in Appendix B.


\section{Conclusion}\label{sec:conclude}

In this paper, we studied QNMs from the Dirac perturbation around the black hole in the dRGT massive gravity, called the dRGT black hole. For the dRGT black hole, a part of the graviton mass can play the role of the cosmological constant so that the dRGT black hole solution is asymptotically AdS/dS, as expressed in Eq.~\eqref{f sol} and Eq.~\eqref{fdRGT}. Conveniently, we characterized the effect of the graviton mass by two parameters, $c_2$ and $\beta_m$. The parameter $c_2$ characterizes the strength of the graviton mass and $\beta_m$ ($0<\beta_m<1$) determines the existence of the horizons. For the Dirac perturbation, we derived the radial Schr\"{o}dinger-like equation and two effective potentials $V_{\pm}$ were obtained. These are super-symmetric partner potentials, so that both give the same spectrum of QNMs. Therefore, we specified the form of $V_+$ in order to obtain the QNMs in this work. The shape of the effective potential is crucially described by three parameters; $c_2$ and $\beta_m$ characterize the graviton mass and $\lambda$ is the eigenvalue of the angular part of the perturbation equations corresponding to angular momentum. We restricted our consideration to the asymptotically dS spacetime so that the potential is barrier-like, with $c_2<0$ and $ 0<\beta_m<1$. This allowed us to use the WKB method to calculate the QNMs.

We first analyzed the behavior of QNMs by considering how the QNM frequency changes according to the change in shape of the potential. We found that the behavior of the frequency is similar to that in quantum mechanics. The higher potential, the faster the wave decay (the higher of the absolute value of the imaginary part of the frequency). In terms of the graviton mass parameters, the higher potential corresponds to lower values of $\beta_m$ and $c_2$. For the angular momentum parameter $\lambda$, the higher potential corresponds to larger values of $\lambda$. This behavior can be roughly seen from Eq.~\eqref{3rd WKB freq}. In the second part of our study, numerical calculations were performed, which verified the veracity of our analytic results. For the numerical calculations, the QMNs were evaluated up to the 3rd and 6th orders of the WKB approximation. We also checked our numerical calculations by using another approximation, namely, the Pad$\acute{e}$ approximation up to the 6th and 13th orders. We found that all results are in good agreement.

Since the dRGT black hole solution is asymptotically dS, we also compared our results to the Schwarzschild-dS black hole in order to distinguish between them. The Dirac QNMs for the Schwarzschild-dS located approximately in a part of the results from the dRGT black hole as shown in Fig.~\ref{fig:QNM3} as well as in Tables \ref{Tab1}, \ref{Tab2}, \ref{Tab3}, \ref{Tab4}, \ref{Tab5} and \ref{Tab6} in Appendix A. In other words, the QNM frequencies for the dRGT black hole can be more or less than ones for Schwarzschild-dS depending on the graviton mass parameters. Actually, for the dRGT black hole, it is possible to obtain faster or slower decay rates of the wave compared to the Schwarzschild-dS black hole. This is due to the fact that the dRGT solution has more free parameters than the single free parameter of the Schwarzschild-dS solution. This provides us one possible way to distinguish/test this kind of modified gravity theory. For example, it is possible to check how the QNMs during the ringdown phase of the black hole mergers deviate from those in the Schwarzschild solution \cite{Card2019}.

It is worthwhile to compare our results with the black hole solutions in other kinds of massive gravity theories which can be found in \cite{fern2015}, for example. However, the black hole solution in \cite{fern2015} is asymptotically flat which is very different from the dRGT solution. Therefore, the potential is crucially different, even though it is also a barrier-like potential. Nevertheless, it is possible to apply our analysis to solution in \cite{fern2015}. Actually, the result agrees with the argument that higher potentials make the wave with QNMs decay faster than the lower one.

In order to characterize the dynamics of the black hole, it is possible to consider other quantities such as the transmission probability or greybody factor. These quantities tell us how the wave emitted near by the black hole can propagate from the black hole. Then the properties of the black hole can be identified through the potential in the master equation of the Dirac perturbation. We leave this investigation for future work. Furthermore, the solution with asymptotically AdS spacetime is also interesting. In this case, it is more complicated to calculate the QNMs since the proper specific boundary conditions for the modes must be imposed. Moreover, there are three possible horizons for the dRGT black hole so that the boundary conditions should be carefully specified. Even though it might be complicated to perform the QNM analysis in this case, it is interesting in the context of the AdS/CFT correspondence. Perhaps, the QNM frequencies or graviton mass may correspond to some quantities in the dual field theory and may provide some imprints of quantum gravity theory. Furthermore, the perturbations from the other fermionic test fields are also of interest, for example, massive gravitino perturbations.

\appendix
\section{The QNM results}
We have explicitly listed the low-lying QNMs corresponding to the content in Sec. \uppercase\expandafter{\romannumeral4} with 3rd and 6th order WKB approximations, and 6th and 13th order revised WKB approach with Pad$\acute{e}$ approximation.

\begin{table}[!htbp]
\centering
\caption{Dirac QNM frequencies for different values of $\beta_{m}$ \\with $\tilde{M}=\alpha_{g}=1$, $c_{2}=-\frac{0.01}{3}$, $l=0$ and $n=0$.}
\begin{tabular}{| c | c | c | c | c| }
\hline
 & \multicolumn{4}{|c|}{QNMs} \\
\hline
$\beta_{m}$ & 3rd WKB & 6th WKB  & Revised 6th WKB & Revised 13th WKB \\
\hline
0.5   & 0.38649 - 0.32936 i & 0.40546 - 0.30224 i & 0.411769 - 0.306595 i & 0.412062 - 0.307285 i\\
0.6   & 0.28083 - 0.20141 i & 0.29345 - 0.18693 i & 0.295206 - 0.190809 i & 0.297504 - 0.191988 i\\
0.7   & 0.20399 - 0.12525 i & 0.21196 - 0.11793 i & 0.212123 - 0.120226 i & 0.212923 - 0.120781 i\\
0.8   & 0.14320 - 0.07566 i & 0.14768 - 0.07246 i & 0.147589 - 0.073619 i & 0.147665 - 0.073817 i\\
0.85  & 0.11592 - 0.05671 i & 0.11900 - 0.05483 i & 0.118880 - 0.055583 i & 0.118886 - 0.055649 i\\
0.9   & 0.08894 - 0.04015 i & 0.09081 - 0.03923 i & 0.090714 - 0.039623 i & 0.090719 - 0.039635 i\\
0.95  & 0.05945 - 0.02462 i & 0.06028 - 0.02434 i & 0.060255 - 0.024468 i & 0.060237 - 0.024454 i\\
\hline
$\Lambda$ & \multicolumn{4}{|c|}{QNMs in Schwarzschild dS $l=0$ and $n=0$.} \\
\hline
0.01  & 0.16917 - 0.09501 i & 0.17503 - 0.09028 i & 0.175008 - 0.091818 i & 0.175258 - 0.092259 i\\
\hline
\end{tabular}
\label{Tab1}
\end{table}

\begin{table}[!htbp]
\centering
\caption{Dirac QNM frequencies for different values of $\beta_{m}$ \\with $\tilde{M}=\alpha_{g}=1$, $c_{2}=-\frac{0.01}{3}$, $l=1$ and $n=0$.}
\begin{tabular}{| c | c | c | c | c | }
\hline
 & \multicolumn{4}{|c|}{QNMs}  \\
\hline
$\beta_{m}$ & 3rd WKB & 6th WKB  & Revised 6th WKB & Revised 13th WKB\\
\hline
0.5   & 0.87844 - 0.30598 i & 0.88848 - 0.30278 i & 0.887859 - 0.304128 i & 0.887831 - 0.304135 i\\
0.6   & 0.62074 - 0.19053 i & 0.62533 - 0.18952 i & 0.625124 - 0.189894 i & 0.625114 - 0.189891 i\\
0.7   & 0.44074 - 0.12043 i & 0.44282 - 0.12012 i & 0.442772 - 0.120208 i & 0.442768 - 0.120206 i\\
0.8   & 0.30304 - 0.07386 i & 0.30391 - 0.07377 i & 0.303912 - 0.073781 i & 0.303905 - 0.073786 i\\
0.85  & 0.24273 - 0.05577 i & 0.24325 - 0.05573 i & 0.243257 - 0.055733 i & 0.243254 - 0.055734 i\\
0.9   & 0.18411 - 0.03977 i & 0.18439 - 0.03975 i & 0.184397 - 0.039751 i & 0.184396 - 0.039751 i\\
0.95  & 0.12143 - 0.02454 i & 0.12153 - 0.02453 i & 0.121537 - 0.024539 i & 0.121537 - 0.024538 i\\
\hline
$\Lambda$ & \multicolumn{4}{|c|}{QNMs in Schwarzschild dS $l=1$ and $n=0$.} \\
\hline
0.01  & 0.36165 - 0.09200 i & 0.36296 - 0.09184 i & 0.362950 - 0.091860 i & 0.362935 - 0.091876 i\\
\hline
\end{tabular}
\label{Tab2}
\end{table}

\begin{table}[!htbp]
\centering
\caption{Dirac QNM frequencies for different values of $\beta_{m}$ \\with $\tilde{M}=\alpha_{g}=1$, $c_{2}=-\frac{0.01}{3}$, $l=1$ and $n=1$.}
\begin{tabular}{| c | c | c | c | c | }
\hline
 & \multicolumn{4}{|c|}{QNMs }  \\
\hline
$\beta_{m}$ & 3rd WKB & 6th WKB  & Revised 6th WKB & Revised 13th WKB\\
\hline
0.5   & 0.78589 - 0.96400 i & 0.79472 - 0.95088 i & 0.793818 - 0.953851 i & 0.795143 - 0.955411 i\\
0.6   & 0.56695 - 0.59479 i & 0.57241 - 0.58923 i & 0.571799 - 0.590105 i & 0.572419 - 0.590775 i\\
0.7   & 0.41015 - 0.37251 i & 0.41317 - 0.37023 i & 0.412984 - 0.370286 i & 0.413156 - 0.370659 i\\
0.8   & 0.28727 - 0.22626 i & 0.28875 - 0.22536 i & 0.288748 - 0.225362 i & 0.288752 - 0.225457 i\\
0.85  & 0.23242 - 0.16996 i & 0.23339 - 0.16940 i & 0.233393 - 0.169405 i & 0.233392 - 0.169443 i\\
0.9   & 0.17829 - 0.12052 i & 0.17889 - 0.12018 i & 0.178895 - 0.120184 i & 0.178895 - 0.120197 i\\
0.95  & 0.11921 - 0.07396 i & 0.11950 - 0.07381 i & 0.119508 - 0.073813 i & 0.119508 - 0.073814 i\\
\hline
$\Lambda$ & \multicolumn{4}{|c|}{QNMs in Schwarzschild dS $l=1$ and $n=1$.} \\
\hline
0.01  & 0.33934 - 0.28358 i & 0.34140 - 0.28225 i & 0.341354 - 0.282252 i & 0.341391 - 0.282449 i \\
\hline
\end{tabular}
\label{Tab3}
\end{table}

\begin{table}[!htbp]
\centering
\caption{Dirac QNM frequencies for different values of $\beta_{m}$ \\with $\tilde{M}=\alpha_{g}=1$, $c_{2}=-\frac{0.01}{3}$, $l=2$ and $n=0$.}
\begin{tabular}{| c | c | c | c | c | }
\hline
 & \multicolumn{4}{|c|}{QNMs }  \\
\hline
$\beta_{m}$ & 3rd WKB & 6th WKB  & Revised 6th WKB & Revised 13th WKB\\
\hline
0.5   & 1.34510 - 0.30386 i & 1.34827 - 0.30360 i & 1.348260 - 0.303613 i & 1.348240 - 0.303624 i\\
0.6   & 0.94532 - 0.18976 i & 0.94668 - 0.18968 i & 0.946688 - 0.189683 i & 0.946684 - 0.189683 i\\
0.7   & 0.66844 - 0.12017 i & 0.66903 - 0.12014 i & 0.669039 - 0.120141 i & 0.669039 - 0.120141 i\\
0.8   & 0.45802 - 0.07380 i & 0.45826 - 0.07379 i & 0.458266 - 0.073788 i & 0.458265 - 0.073789 i\\
0.85  & 0.36627 - 0.05575 i & 0.36641 - 0.05575 i & 0.366416 - 0.055752 i & 0.366415 - 0.055752 i\\
0.9   & 0.27735 - 0.03977 i & 0.27743 - 0.03977 i & 0.277432 - 0.039773 i & 0.277432 - 0.039773 i\\
0.95  & 0.18258 - 0.02455 i & 0.18261 - 0.02455 i & 0.182610 - 0.024553 i & 0.182610 - 0.024553 i\\
\hline
$\Lambda$ & \multicolumn{4}{|c|}{QNMs in Schwarzschild dS $l=2$ and $n=0$.} \\
\hline
0.01  & 0.54755 - 0.09184 i & 0.54792 - 0.09183 i & 0.547926 - 0.091829 i & 0.547926 - 0.091829 i \\
\hline
\end{tabular}
\label{Tab4}
\end{table}

\begin{table}[!htbp]
\centering
\caption{Dirac QNM frequencies for different values of $\beta_{m}$ \\with $\tilde{M}=\alpha_{g}=1$, $c_{2}=-\frac{0.01}{3}$, $l=2$ and $n=1$.}
\begin{tabular}{| c | c | c | c | c |}
\hline
 & \multicolumn{4}{|c|}{QNMs }  \\
\hline
$\beta_{m}$ & 3rd WKB & 6th WKB  & Revised 6th WKB & Revised 13th WKB\\
\hline
0.5   &1.27454 - 0.93365 i& 1.28008 - 0.93084 i  & 1.280070 - 0.930835 i & 1.280040 - 0.931082 i\\
0.6   &0.90631 - 0.57976 i& 0.90890 - 0.57869 i  & 0.908899 - 0.578694 i & 0.908892 - 0.578733 i\\
0.7   &0.64708 - 0.36536 i& 0.64828 - 0.36493 i  & 0.648284 - 0.364932 i & 0.648285 - 0.364938 i\\
0.8   &0.44737 - 0.22334 i& 0.44790 - 0.22316 i  & 0.447900 - 0.223162 i & 0.447902 - 0.223164 i\\
0.85  &0.35942 - 0.16834 i& 0.35975 - 0.16822 i  & 0.359757 - 0.168227 i & 0.359758 - 0.168228 i\\
0.9   &0.27356 - 0.11979 i& 0.27376 - 0.11972 i  & 0.273763 - 0.119726 i & 0.273763 - 0.119726 i\\
0.95  &0.18117 - 0.07377 i& 0.18126 - 0.07374 i  & 0.181264 - 0.073747 i & 0.181264 - 0.073746 i\\
\hline
$\Lambda$ & \multicolumn{4}{|c|}{QNMs in Schwarzschild dS $l=2$ and $n=1$} \\
\hline
0.01  &0.53216 - 0.27879 i& 0.53292 - 0.27855 i  & 0.532920 - 0.278548 i & 0.532921 - 0.278550 i \\
\hline
\end{tabular}
\label{Tab5}
\end{table}

\begin{table}[!htbp]
\centering
\caption{Dirac QNM frequencies for different values of $\beta_{m}$ \\with $\tilde{M}=\alpha_{g}=1$, $c_{2}=-\frac{0.01}{3}$, $l=2$ and $n=2$.}
\begin{tabular}{| c | c | c | c | c | }
\hline
 & \multicolumn{4}{|c|}{QNMs }  \\
\hline
$\beta_{m}$ & 3rd WKB & 6th WKB  & Revised 6th WKB & Revised 13th WKB\\
\hline
0.5   &1.16848 - 1.59407 i& 1.16882 - 1.61096 i  & 1.168210 - 1.610720 i & 1.169640 - 1.610490 i\\
0.6   &0.84478 - 0.98643 i& 0.84387 - 0.99430 i  & 0.843780 - 0.994246 i & 0.844222 - 0.994234 i\\
0.7   &0.61182 - 0.61916 i& 0.61088 - 0.62271 i  & 0.610871 - 0.622698 i & 0.611053 - 0.622695 i\\
0.8   &0.42894 - 0.37665 i& 0.42833 - 0.37797 i  & 0.428335 - 0.377966 i & 0.428412 - 0.377987 i\\
0.85  &0.34725 - 0.28308 i& 0.34687 - 0.28371 i  & 0.346875 - 0.283709 i & 0.346918 - 0.283734 i\\
0.9   &0.26662 - 0.20079 i& 0.26649 - 0.20097 i  & 0.266492 - 0.200969 i & 0.266504 - 0.200985 i\\
0.95  &0.17850 - 0.12325 i& 0.17854 - 0.12322 i  & 0.178551 - 0.123218 i & 0.178550 - 0.123220 i\\
\hline
$\Lambda$ & \multicolumn{4}{|c|}{QNMs in Schwarzschild dS $l=2$ and $n=2$} \\
\hline
0.01  &0.50629 - 0.47188 i& 0.50552 - 0.47412 i & 0.505521 - 0.474119 i & 0.505616 - 0.474121 i \\
\hline
\end{tabular}
\label{Tab6}
\end{table}

\begin{table}[!htbp]
\centering
\caption{Dirac QNM frequencies for different values of $c_{2}$ \\with $\tilde{M}=\alpha_{g}=1$, $\beta_{m}=0.8$, $l=0$ and $n=0$.}
\begin{tabular}{| c | c | c | c | c | }
\hline
 & \multicolumn{4}{|c|}{QNMs }  \\
\hline
$c_{2}$ & 3rd WKB & 6th WKB  & Revised 6th WKB & Revised 13th WKB\\
\hline
-0.02/3  &0.19826 - 0.12154 i & 0.20603 - 0.11454 i  & 0.206167 - 0.116688 i & 0.206861 - 0.117322 i\\
-0.03/3  &0.23922 - 0.16062 i & 0.24967 - 0.15001 i  & 0.250315 - 0.153272 i & 0.251764 - 0.153657 i\\
-0.04/3  &0.27296 - 0.19588 i & 0.28565 - 0.18181 i  & 0.287128 - 0.185903 i & 0.289063 - 0.186408 i \\
-0.05/3  &0.30214 - 0.22856 i & 0.31677 - 0.21119 i  & 0.319255 - 0.215908 i & 0.321927 - 0.216762 i \\
-0.06/3  &0.37548 - 0.31714 i & 0.34446 - 0.23880 i  & 0.348049 - 0.243965 i & 0.351207 - 0.243686 i\\
\hline
\end{tabular}
\label{Tab7}
\end{table}

\begin{table}[!htbp]
\centering
\caption{Dirac QNM frequencies for different values of $c_{2}$ \\with $\tilde{M}=\alpha_{g}=1$, $\beta_{m}=0.8$, $l=1$ and $n=0$.}
\begin{tabular}{| c | c | c | c | c | }
\hline
 & \multicolumn{4}{|c|}{QNMs }  \\
\hline
$c_{2}$ & 3rd WKB & 6th WKB  & Revised 6th WKB & Revised 13th WKB\\
\hline
-0.02/3 &0.42681 - 0.11734 i  & 0.42877 - 0.11705 i  & 0.428734 - 0.117132 i & 0.428726 - 0.117129 i\\
-0.03/3 &0.52115 - 0.15388 i  & 0.52428 - 0.15330 i  & 0.524184 - 0.153489 i & 0.524170 - 0.153482 i\\
-0.04/3 &0.60028 - 0.18655 i  & 0.60461 - 0.18560 i  & 0.604444 - 0.185940 i & 0.604423 - 0.185931 i \\
-0.05/3 &0.66969 - 0.21662 i  & 0.67525 - 0.21525 i  & 0.675006 - 0.215767 i & 0.674976 - 0.215755 i\\
-0.06/3 &0.73220 - 0.24477 i  & 0.73902 - 0.24293 i  & 0.738683 - 0.243654 i & 0.738644 - 0.243642 i\\
\hline
\end{tabular}
\label{Tab8}
\end{table}

\begin{table}[!htbp]
\centering
\caption{Dirac QNM frequencies for different values of $c_{2}$ \\with $\tilde{M}=\alpha_{g}=1$, $\beta_{m}=0.8$, $l=1$ and $n=1$.}
\begin{tabular}{| c | c | c | c | c | }
\hline
 & \multicolumn{4}{|c|}{QNMs }  \\
\hline
$c_{2}$ & 3rd WKB & 6th WKB  & Revised 6th WKB & Revised 13th WKB\\
\hline
-0.02/3   &0.39922 - 0.36147 i & 0.40209 - 0.35921 i  & 0.402063 - 0.359262 i & 0.402099 - 0.359609 i\\
-0.03/3   &0.48308 - 0.47579 i & 0.48717 - 0.47196 i  & 0.487045 - 0.472284 i & 0.487225 - 0.472800 i\\
-0.04/3   &0.55258 - 0.57840 i & 0.55774 - 0.57290 i  & 0.557475 - 0.573569 i & 0.557870 - 0.574263 i\\
-0.05/3   &0.61300 - 0.67314 i & 0.61908 - 0.66589 i  & 0.618793 - 0.666903 i & 0.619328 - 0.667844 i\\
-0.06/3   &0.66701 - 0.76205 i & 0.67390 - 0.75302 i  & 0.673610 - 0.754384 i & 0.674174 - 0.754743 i\\
\hline
\end{tabular}
\label{Tab9}
\end{table}

\begin{table}[!htbp]
\centering
\caption{Dirac QNM frequencies for different values of $c_{2}$ \\with $\tilde{M}=\alpha_{g}=1$, $\beta_{m}=0.8$, $l=2$ and $n=0$.}
\begin{tabular}{| c | c | c | c | c | }
\hline
 & \multicolumn{4}{|c|}{QNMs }  \\
\hline
$c_{2}$ & 3rd WKB & 6th WKB  & Revised 6th WKB & Revised 13th WKB\\
\hline
-0.02/3   &0.64681 - 0.11716 i & 0.64737 - 0.11713 i  & 0.647377 - 0.117132 i & 0.647376 - 0.117132 i\\
-0.03/3   &0.79135 - 0.15354 i & 0.79226 - 0.15348 i  & 0.792264 - 0.153486 i & 0.792262 - 0.153485 i\\
-0.04/3   &0.91298 - 0.18601 i & 0.91426 - 0.18593 i  & 0.914269 - 0.185934 i & 0.914266 - 0.185934 i\\
-0.05/3   &1.01998 - 0.21587 i & 1.02165 - 0.21575 i  & 1.021660 - 0.215756 i & 1.021650 - 0.215756 i\\
-0.06/3   &1.11658 - 0.24379 i & 1.11867 - 0.24363 i  & 1.118670 - 0.243639 i & 1.118660 - 0.243641 i\\
\hline
\end{tabular}
\label{Tab10}
\end{table}

\begin{table}[!htbp]
\centering
\caption{Dirac QNM frequencies for different values of $c_{2}$ \\with $\tilde{M}=\alpha_{g}=1$, $\beta_{m}=0.8$, $l=2$ and $n=1$.}
\begin{tabular}{| c | c | c | c | c | }
\hline
 & \multicolumn{4}{|c|}{QNMs }  \\
\hline
$c_{2}$ & 3rd WKB & 6th WKB  & Revised 6th WKB & Revised 13th WKB\\
\hline
-0.02/3  &0.62784 - 0.35544 i & 0.62899 - 0.35499 i  & 0.628997 - 0.354994 i & 0.629000 - 0.355002 i\\
-0.03/3  &0.76477 - 0.46662 i & 0.76658 - 0.46586 i  & 0.766586 - 0.465853 i & 0.766589 - 0.465877 i\\
-0.04/3  &0.87924 - 0.56613 i & 0.88173 - 0.56500 i  & 0.881730 - 0.564996 i & 0.881733 - 0.565041 i\\
-0.05/3  &0.97940 - 0.65778 i & 0.98257 - 0.65626 i  & 0.982569 - 0.656256 i & 0.982570 - 0.656331 i\\
-0.06/3  &1.06942 - 0.74364 i & 1.07327 - 0.74170 i  & 1.073270 - 0.741696 i & 1.073270 - 0.741808 i\\
\hline
\end{tabular}
\label{Tab11}
\end{table}

\begin{table}[!htbp]
\centering
\caption{Dirac QNM frequencies for different values of $c_{2}$ \\with $\tilde{M}=\alpha_{g}=1$, $\beta_{m}=0.8$, $l=2$ and $n=2$.}
\begin{tabular}{| c | c | c | c | c | }
\hline
 & \multicolumn{4}{|c|}{QNMs }  \\
\hline
$c_{2}$ & 3rd WKB & 6th WKB  & Revised 6th WKB & Revised 13th WKB\\
\hline
-0.02/3   &0.59593 - 0.60072 i & 0.59495 - 0.60370 i & 0.594941 - 0.603688 i & 0.595173 - 0.603723 i\\
-0.03/3   &0.72089 - 0.78968 i & 0.71967 - 0.79444 i & 0.719655 - 0.794419 i & 0.720070 - 0.794465 i\\
-0.04/3   &0.82432 - 0.95900 i & 0.82294 - 0.96560 i & 0.822921 - 0.965577 i & 0.823582 - 0.965611 i\\
-0.05/3   &0.91410 - 1.11509 i & 0.91263 - 1.12359 i & 0.912598 - 1.123560 i & 0.913527 - 1.123580 i\\
-0.06/3   &0.99423 - 1.26143 i & 0.99273 - 1.27185 i & 0.992690 - 1.271810 i & 0.993932 - 1.271790 i\\
\hline
\end{tabular}
\label{Tab12}
\end{table}

\FloatBarrier

\section{Numerical precision of the QNM results from different approximation}
As mentioned in Sec.~\ref{sec:QNM}, we evaluated our data with the WKB method and the revised WKB method with Pad$\acute{e}$ approximation using the Wolfram Mathematica with versions $8.0$ to $11.0$. Some minimal differences might happen when calculations are made with different computing platforms or different versions of Wolfram Mathematica. In order to obtain reproducible data, we have to assign the precision numbers step by step in the calculation.

In this section, we took a specific case as an example to show the QNMs with different settings of the numerical precision. In Table~\ref{Tab13}, the ``Initial parameters" denotes the initial precision values of the coefficients $c_{0}$ and $c_{1}$, and the ``For the methods" denotes the assigned precision to evaluate QNM frequencies using the WKB or revised WKB method with Pad$\acute{e}$ approximation. For the fourth column, we set the precision of initial parameters assigned from the Wolfram Mathematica and set 50 digits precision for the revised WKB method to evaluate QMN frequencies. We found that the results are successfully evaluated up to the 10th order as inferred from the frequencies with black color  in the fourth column. Note that the frequencies with blue color and ones with red color denote the unsuccessful evaluation with warning and error messages, respectively, on the precision problem in the calculation. 

For other columns of revised WKB method, the frequencies are evaluated with increasing precisions. We found that it is sufficient to evaluate QNM frequencies up to 13th order by the precision settings as shown in the last column.

\begin{table}[!htbp]
\tiny
\centering
\caption{Comparison of numerical precision of the Dirac QNM results obtain from WKB and revised WKB methods with $M=\alpha_{g}=1$, $\beta_{m}=0.8$, $l=0$, $n=0$ and $c_{2}=-\frac{0.04}{3}$.}
\begin{tabular}{| c | c | c | c | c | c | c | }
\hline
 & \multicolumn{6}{|c|}{Methods }  \\
\hline
order & WKB          & WKB      & Revised WKB   & Revised WKB &Revised WKB & Revised WKB\\
\hline
 & \multicolumn{6}{|c|}{number of digits of precision setting in Mathematica }  \\
\hline
Initial parameter  & non   & non & non   &  500  &  1000    & 1000\\
\hline
For the methods    & non   & 50  & 50    &  50   &   50     & 100\\
\hline
\hline
order & \multicolumn{6}{|c|}{QNMs results}  \\
\hline
1th   &0.40053 - 0.18145 i &0.40053 - 0.18145 i & 0.33232 - 0.15055 i & 0.33232 - 0.15055 i& 0.33232 - 0.15055 i & 0.33232 - 0.15055 i\\
2nd   &0.30489 - 0.23836 i &0.30489 - 0.23836 i & 0.28416 - 0.17889 i & 0.28416 - 0.17889 i& 0.28416 - 0.17889 i & 0.28416 - 0.17889 i\\
3rd   &0.27296 - 0.19588 i &0.27296 - 0.19588 i & 0.28664 - 0.18763 i & 0.28664 - 0.18763 i& 0.28664 - 0.18763 i & 0.28664 - 0.18763 i\\
4th   &0.28350 - 0.18860 i &0.28349 - 0.18860 i & 0.28708 - 0.18617 i & 0.28708 - 0.18617 i& 0.28708 - 0.18617 i & 0.28708 - 0.18617 i\\
5th   &0.28230 - 0.18679 i &0.28099 - 0.18482 i & 0.28766 - 0.18596 i & 0.28766 - 0.18596 i& 0.28766 - 0.18596 i & 0.28766 - 0.18596 i\\
6th   &{\color{red}0.68075 - 0.07746 i} &0.28565 - 0.18181 i & 0.28712 - 0.18590 i & 0.28712 - 0.18590 i& 0.28712 - 0.18590 i & 0.28712 - 0.18590 i\\
7th   &                    &                    & 0.28685 - 0.18369 i & 0.28685 - 0.18369 i& 0.28685 - 0.18369 i & 0.28685 - 0.18369 i\\
8th   &                    &                    & 0.28961 - 0.18639 i & {\color{blue}0.28965 - 0.18608 i}& {\color{blue}0.28965 - 0.18608 i} & 0.28961 - 0.18639 i\\
9th   &                    &                    & 0.28757 - 0.18815 i & {\color{red}0.28668 - 0.18835 i}& {\color{blue}0.28668 - 0.18835 i} & 0.28757 - 0.18815 i\\
10th  &                    &                    & 0.28807 - 0.18628 i & {\color{red}0.29030 - 0.18455 i}& {\color{red}0.29030 - 0.18455 i} & 0.28807 - 0.18628 i\\
11th  &                    &                    & {\color{blue}0.28781 - 0.18564 i} & {\color{red}0.29369 - 0.17284  i}& {\color{red}0.29369 - 0.17284 i} & 0.28781 - 0.18564 i\\
12th  &                    &                    & {\color{blue}0.28891 - 0.18591 i} & {\color{red}0.29384 - 0.17303 i}& {\color{red}0.29384 - 0.17303 i} & 0.28891 - 0.18591 i\\
13th  &                    &                    & {\color{red}0.32046 - 0.16385 i}  & {\color{red}0.29307 - 0.172609 i}& {\color{red}0.29307 - 0.17260 i} & 0.28906 - 0.18640 i\\
\hline
{\color{red}Red words}& \multicolumn{6}{|c|}{Including error messages in the calculation}  \\
\hline
{\color{blue}Blue words}& \multicolumn{6}{|c|}{Including some warning messages on the precision problem in the calculation}  \\
\hline
\end{tabular}
\label{Tab13}
\end{table}

\section*{Acknowledgement}

We would like to thank Prof. Matthew James Lake and Dr. Patharadanai Nuchino for reading through the manuscript and correcting some grammatical error. PW is supported by the Thailand Research Fund (TRF) through grant no. MRG6180003. PW also would like to thank the Department of Mathematics and Computer Science, Faculty of Science, Chulalongkorn University for hospitality while this work was in progress and SERB-DST, India for the ASEAN project IMRC/AISTDF/CRD/2018/000042.

\end{document}